\documentclass{emulateapj}

\def\kpc{\,{\rm kpc}}
\def\om{\Omega_{\rm p}}
\def\etal{{et al.}}
\def\eg{{\it e.g.}}

\def\ie{{\it i.e.}}

\def\rd{R_{\rm d}}
\def\kms{$\mathrm {km}~\mathrm{s}^{-1}$}

\def\kmsM{$\mathrm {km}~\mathrm{s}^{-1}~ \mathrm{Mpc}^{-1}$}
\def\degrees{^\circ}
\def\spose#1{\hbox to 0pt{#1\hss}}
\def\gtsim{\mathrel{\spose{\lower.5ex \hbox{$\mathchar"218$}}
     \raise.4ex\hbox{$\mathchar"13E$}}}
\def\ltsim{\mathrel{\spose{\lower.5ex\hbox{$\mathchar"218$}}
     \raise.4ex\hbox{$\mathchar"13C$}}}

\slugcomment{Draft to ApJ}
\lefthead{Debattista \etal}
\righthead{The Secular Evolution of Disk Structural Parameters}

\begin{document}

\title{The Secular Evolution of Disk Structural Parameters}

\author{Victor P.~Debattista\altaffilmark{1,}\altaffilmark{2}}
\affil{Institut f\"ur Astronomie, ETH Z\"urich, CH-8093,
  Z\"urich, Switzerland}
 \email{debattis@astro.washington.edu}

\author{Lucio Mayer}
\affil{Institut f\"ur Astronomie, ETH Z\"urich, CH-8093,
  Z\"urich, Switzerland}
\email{lucio@phys.ethz.ch}

\author{C. Marcella Carollo}
\affil{Institut f\"ur Astronomie, ETH Z\"urich, CH-8093,
  Z\"urich, Switzerland}
\email{marcella.carollo@phys.ethz.ch}

\author{Ben Moore}
\affil{Department of Theoretical Physics, University of Z\"urich,
  Winterthurerstrasse 190, CH-8057, Z\"urich, Switzerland}
\email{moore@physik.unizh.ch}

\author{James Wadsley} \affil{Department of Physics and Astronomy,
  McMaster University, 1280 Main St. West, Hamilton, Ontario L8S 4M1,
  Canada} 
\email{wadsley@mcmaster.ca}

\author{Thomas Quinn} \affil{Astronomy Department, University of
  Washington, Box 351580, Seattle WA, 98195, USA}
\email{trq@astro.washington.edu}

\altaffiltext{1}{current address: Astronomy Department, University of
Washington, Box 351580, Seattle WA, 98195, USA}
\altaffiltext{2}{Brooks Prize Fellow}
 
\begin{abstract}
We present a comprehensive series of simulations to study the secular
evolution of disk galaxies expected in a LCDM universe.  Our
simulations are organized in a hierarchy of increasing complexity,
ranging from rigid-halo collisionless simulations to fully live
simulations with gas and star formation.

Our goal is to examine which structural properties of disk galaxies
may result from secular evolution rather than from hierarchical
assembly.  In the vertical direction, we find that various mechanisms
lead to heating, the strongest of which is the buckling instability of
a bar, which leads to peanut-shaped bulges; these can be recognized
face-on even in the presence of gas.
We find that bars are robust structures that survive buckling and
require a large ($\sim 20\%$ of the total mass of the disk) central
mass concentration to be destroyed. This can occur in dissipative
simulations, where bars induce strong gas inflows, but requires that
radiative cooling overcome heating.
We show how angular momentum redistribution leads to increasing
central densities and disk scale lengths and to profile breaks at
large radii. The breaks in these simulations are in excellent
agreement with observations, even when the evolution is collisionless.
Disk scale-lengths increase even when the total disk angular momentum
is conserved; thus mapping halo angular momenta to scale-lengths is
non-trivial.  A decomposition of the resulting profile into a
bulge$+$disk gives structural parameters in reasonable agreement with
observations although kinematics betray their bar nature.

These findings have important implications for galaxy formation
models, which have so far ignored or introduced in a very simplified
way the effects of non-axisymmetric instabilities on the morphological
evolution of disk galaxies.
\end{abstract}


\section{Introduction}

In the current paradigm, galaxy formation is hierarchical (\eg\ White
\& Rees 1978; Steinmetz \& Navarro 2002).  Indeed, evidence can be
found of continued accretion onto both the Milky Way (Ibata \etal\
1994; Helmi \etal\ 1999) and M31 (Ferguson \etal\ 2002).  Although the
standard framework of disk formation in such cosmogonies was
formulated some time ago (White \& Rees 1978; Fall \& Efstathiou
1980), disk galaxy formation remains a challenging problem for both
simulations (\eg\ Navarro \& Steinmetz 2000; Abadi \etal\ 2003) and
semi-analytical models (\eg\ Somerville \& Primack 1999; Hatton \etal\
2003; van den Bosch 1998, 2000, 2001, 2002; Firmani \& Avila-Reese
2000; Mo \etal\ 1998; Dalcanton \etal\ 1997; Benson \etal\ 2003; Cole
\etal\ 1994, 2000; Baugh \etal\ 1996; Kauffmann \etal\ 1993).  The
latest generation of simulations produces disks having roughly the
correct sizes and structural properties (Governato \etal\ 2004; Abadi
\etal\ 2003; Sommer-Larsen \etal\ 2003). This is due partly to the
increased resolution that reduces the artificial loss of angular
momentum of the baryonic component.  At the same time, it has been
realized that only halos with a quiet merging history after $z\sim 2$
can host disk galaxies today (but see Springel \& Hernquist 2005).
Therefore in the current picture most of the mass of the disk is
assembled from the smooth accretion of gas cooling inside the dark
halo (combined with the accretion along cold filaments for lower mass
objects; Katz \& Gunn 1991; Katz \etal\ 2003; Birnboim \& Dekel 2003;
Kere\v{s} \etal\ 2005) following an intense phase of merger activity
that can give rise to the stellar halo and a massive old bulge.

During disk assembly, secular evolution must have played a role in
shaping the structure of disk galaxies as we see them at
$z=0$. Non-axisymmetric instabilities, particularly bars, drive a
substantial redistribution of mass and angular momentum in the disk. A
possible product of bar-driven evolution is the formation of a
bulge-like component. Such a component will have a stellar population
similar to that of the disk and is thus younger than the old spheroid
formed by the last major merger.  Low-mass bulges may result from such
a mechanism; observed low-mass bulges have disk-like,
almost-exponential stellar density (Andredakis \& Sanders 1994;
Courteau \etal\ 1996; de Jong 1996; Carollo \etal\ 2002; Carollo
\etal\ 2001; Carollo 1999; Carollo \etal\ 1998; Carollo \etal\ 1997;
MacArthur \etal\ 2003) and in some cases disk-like, cold kinematics
(Kormendy 1993; Kormendy \etal\ 2002).  Comparison between bulge and
disk parameters shows a correlation between the scale-lengths of
bulges and disks (de Jong 1996; MacArthur \etal\ 2003) and, on
average, similar colors in bulges and inner disks (Terndrup \etal\
1994; Peletier \& Balcells 1996; Courteau \etal\ 1996; Carollo \etal\
2001).  The disk-like properties of bulges and the links between bulge
and disk properties have been suggested to indicate that bulges may
form through the evolution of disk dynamical instabilities such as
bars, which are present in about $70\%$ of disk galaxies (Knapen 1999;
Eskridge \etal\ 2000).  Further evidence for secular evolution comes
from edge-on galaxies, where bulges are often found to be box- or
peanut-shaped (L\"utticke \etal\ 2000), a shape associated with
evolution driven by the presence of a bar (Combes \& Sanders 1981;
Pfenniger 1984; Combes \etal\ 1990; Pfenniger \& Friedli 1991; Raha
\etal\ 1991; Kuijken \& Merrifield 1995; Bureau \& Freeman 1999;
Bureau \& Athanassoula 1999; Athanassoula \& Bureau 1999).

Secular evolution must also be considered for comparisons of
predictions based on semi-analytic models of disk formation and
observations to be meaningful.  In the standard picture baryons
cooling in dark halos to form a disk have the same specific angular
momentum of the dark matter, and this is conserved during
collapse. Then the distribution of disk scale-lengths can be computed
from the known distribution of halo angular momenta (\eg\ Dalcanton
\etal\ 1997; Mo \etal\ 1998; van den Bosch 1998).  However, de Jong \&
Lacey (2000) found that the width of the observed disk scale-length
distribution at fixed luminosity is smaller than that predicted by
such simple models. This suggests that the mapping between initial
halo angular momenta and disk scale-lengths cannot be so trivial.  In
addition to complications from the cosmological side, including that
the initial specific angular momentum distribution of the baryons need
not be like that of the dark halo (\eg\ van den Bosch et al. 2002) and
angular momentum distributions favoring disks more centrally
concentrated than exponential (Firmani \& Avila-Reese 2000; van den
Bosch 2001; Bullock \etal\ 2001), secular evolution will also change
disk structure.  In the simplest prescriptions this is not considered,
while it is known since Hohl (1971) that bars can drive substantial
evolution of disk profiles.

Which structural properties of present-day disk galaxies are
primordial and which are the result of internal evolution?  Already by
$z\sim 1$ a population of disk galaxies with scale-lengths similar to
those of a local population is observed (Lilly \etal\ 1998),
suggesting that the structural properties of disk galaxies have not
changed substantially since then. If the quiescent phase of disk
assembly starts early, as current cosmological simulations suggest,
secular evolution might have already been operating by
$z=1$. Verification that the secular evolutionary timescale can be
sufficiently short is thus required.  With the present algorithms and
computing power, simulations of individual isolated galaxies are best
able to achieve sufficient force and mass resolution to address such
issues.  Simulations of individual galaxies decoupled from the
hierarchical growth have the additional advantage of exploring
directly the role that internal secular evolution plays in shaping the
currently observed galaxy population.

In this paper we report on a series of such simulations exploring how
disks evolve in the presence of a bar.  An important goal of this
paper is to demonstrate the wealth of structural properties possible
in cosmologically-motivated disk galaxies and to identify what
properties of the mass distribution of disk galaxies may result from
internal evolution, rather than arising directly from hierarchical
assembly.  We identify several structural properties we would like to
test secular evolution for: vertical thickening, inner profile
steepening, profile breaks, and bar destruction.  We explore
simulations in which the full disk-halo interaction is self-consistent
in cosmologically-motivated dark matter halos with and without gas and
star formation.  We use various prescriptions for gas physics, to
understand how this impacts the evolution.  We also use rigid-halo
simulations as in Debattista \etal\ (2004; hereafter Paper I) which
allow us to study a variety of phenomena at high spatial and mass
resolution.  These simulations also help us to isolate physical
mechanisms by which secular evolution occurs.  Together these
simulations allow us to assess the impact of secular evolution on disk
galaxies.

%%%%%%%%%%%%%%%%%%%%%%%%%%%%%%%%%%%%%%%%%%%%%%%%%%%%%%%%%%%%%%%%%%%%%%

\section{Methods}
\label{sec:methods}

\subsection{Live-halo models}
\label{ssec:livehalos}

\begin{table*}[!ht]
\begin{centering}
\begin{tabular}{ccccc|ccc}\hline 
\multicolumn{1}{c}{Run} &
\multicolumn{1}{c}{$M_{disk} [10^{10} M_{\odot}]$} &
\multicolumn{1}{c}{$R_{\rm d} [kpc]$} &
\multicolumn{1}{c}{$Q_{star}$} &
\multicolumn{1}{c}{$f_d$} &
\multicolumn{1}{c}{$f_{\rm gas}$} &
\multicolumn{1}{c}{$Q_{gas}$} &
\multicolumn{1}{c}{Gas physics} \\ \hline 
%                                             
% lucio2 
NC1 & 4.92  &  3.55 & 0.7 & 0.05 & 0.0 &  -  & -   \\ 
% 		   			       
% lucio1 
NC2 & 4.92 & 3.55 & 1.0 & 0.05 & 0.0 &  -  & -   \\ 
% 		   			       
% lucio3 
NC3 & 4.92 & 3.55 & 1.7 & 0.05 & 0.0 &  -  & -   \\ 
% 		   			       
% lucio4 
NG1 & 5.45 & 3.44 & 1.7 & 0.05 & 0.1 &  4 & RC   \\ 
% 		   			       
% lucio7 
NG2 & 5.45 & 3.44 & 1.7 & 0.05 & 0.1 &  4 & A   \\ 
% 		   			       
% lucio8 
NG3 & 5.45 & 3.44 & 1.7 & 0.05 & 0.1 &  4 & RC + SF   \\ 
% 		   			       
% lucio5 
NG4 & 10.9 & 2.35& 1.7 & 0.11 & 0.5 & 0.4 & A  \\ 
% 		   			       
% lucio6 
NG5 & 5.45 & 3.44 & 1.7 & 0.05 & 0.5 & 0.8 & RC   \\ \hline 
\end{tabular}
\caption{The sample of live-halo simulations in this paper.} 
$M_{disk}$, $R_{\rm d}$, $Q_{star}$, $Q_{gas}$ are the disk mass,
initial disk scale-length, and the minimum Toomre-$Q$ parameter of,
respectively, the stellar and the gaseous disk.  $f_d$ is the fraction
of disk (stars+gas) to dark matter mass and $f_{\rm gas}$ is the
fraction of the baryonic mass which is in the gaseous state in the
initial conditions, while ``Gas physics'' lists the physics of the gas
used: 'RC' refers to 'radiative cooling', 'A' refers to 'adiabatic'
and 'SF' refers to 'star formation'.
\label{tab:live}
\end{centering}
\end{table*}

\begin{figure}[!ht]
  \plotone{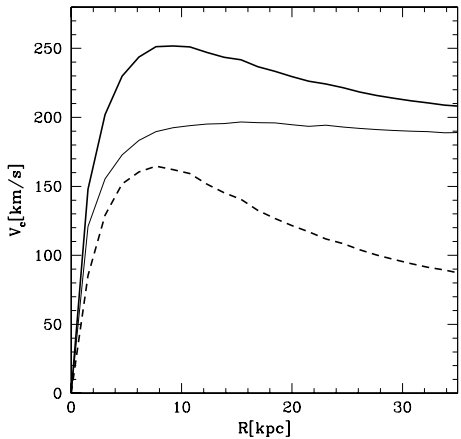}
\caption{Initial rotation curve of the collisionless live-halo
 models. The thick solid line is the total curve, while the thin solid
 and dashed lines represent the separate contributions of,
 respectively, the dark matter and stellar component.
\label{fig:rc1}}
\end{figure}

Live-halo models are built using the technique developed by Hernquist
(1993; see also Springel \& White 1999). The structural properties of
halos and disks are tied together by the scaling relations expected in
the currently favored structure formation model, $\Lambda$CDM.  We
start by choosing the value of the circular velocity of the halo at
the virial radius, $V_{vir}$, which, for an assumed cosmology
(hereafter $\Omega_0=0.3$, $\Lambda=0.7$, $H_0=65$ \kmsM),
automatically determines the virial mass, $M_{vir}$, and virial
radius, $R_{vir}$, of the halo (Mo, Mao \& White 1998).  Halos are
isotropic and have angular momentum that is specified by the spin
parameter, $\lambda = [J^2 |E|/ (G^2 M_{vir}^5)]^{1/2}$, where $J$ and
$E$ are, respectively, the total angular momentum and total energy of
the halo and $G$ is the gravitational constant. We use $\lambda=0.045$
throughout, close to the mean value measured in cosmological
simulations (\eg\ Gardner 2001).  The halo density profile is an NFW
(Navarro, Frenk \& White 1996) with a given value of the concentration
$c$, where $c=R_{vir}/r_s$, $r_s$ being the halo scale radius.  The
higher the concentration, the higher the halo density near the center
at a given value of $M_{vir}$ (and therefore the more steeply rising
is its inner rotation curve).  Adiabatic contraction of the halo due
to the presence of the disk is taken into account by assuming that the
spherical symmetry of the halo is retained and that the angular
momentum of individual dark matter orbits is conserved (see Springel
\& White 1999).  The disk mass fraction relative to the halo virial
mass, $f_d=M_d/M_{vir}$ is $0.05$, consistent with estimates for
galaxies in the local Universe (e.g.  Jimenez, Verde \& Oh, 2003) and
is conservatively lower than the estimate of the universal baryonic
mass fraction yielded by {\it WMAP} (Spergel et al.  2003). Our models
implicitly assume that the disk forms out of collapsed gas that
started with the same specific angular momentum as the halo and that
such angular momentum was conserved during infall (Mo, Mao \& White
1998).  The disk has an exponential surface density profile with scale
length $R_d$ that is determined by the value of $\lambda$ (which sets
the degree of available centrifugal support) and by the values of $c$,
$f_d$, and $M_{vir}$ (which together set the depth of the potential
well).  The setup of the stellar disk is complete once the Toomre
parameter, $Q(R)$, is assigned (Toomre 1964).  This corresponds to
fixing the local radial velocity dispersion $\sigma_R$, since
$Q(R)=\sigma_R \kappa/3.36G \Sigma_s$, where $\kappa$ is the local
epicyclic frequency, $G$ is the gravitational constant, and $\Sigma_s$
is the disk surface density.  The velocity field of the disk is
calculated as in Springel \& White (1999; see also Hernquist 1993); in
particular, the radial and vertical velocity dispersions are assumed
to be equal, and the azimuthal velocity dispersion is determined from
the radial dispersion using the epicyclic approximation.  Dark matter
halos are sampled using $10^6$ particles and stellar disks by $2\times
10^5$ particles.  The gravitational softening of both components is
equal to $300$ pc. We reran a few simulations with higher resolution
($5 \times 10^6$ halo particles and $5\times 10^5$ disk particles) and
smaller softenings ($50$ pc; see Paper I) to check for resolution
effects. We found the analysis presented in this paper to be fairly
insensitive to the resolution adopted, and in the remainder we always
show results for runs with the standard resolution.  The live-halo
models used in this paper have structural parameters in line with the
expectations of $\Lambda$CDM models for a Milky-Way sized system.
These models are similar to the mass models of the Milky Way presented
by Klypin, Zhao \& Somerville (2002).  The rotation curve of the
collisionless models used in the paper is shown in Figure
\ref{fig:rc1}. All collisionless models have the same rotation curve
since they differ only in terms of their Toomre parameter. These
simulations were carried out with the parallel multistepping tree code
PKDGRAV (Stadel 2001).

\subsection{SPH simulations}

Gasdynamical simulations were carried out with GASOLINE, an extension
of PKDGRAV (Stadel 2001) that uses smoothed particle hydrodynamics
(SPH) to solve the hydrodynamical equations (Wadsley, Stadel \& Quinn
2004). The gas is ideal with equation of state $P=(\gamma -1)\rho u$,
where $P$ is the pressure, $\rho$ is the density, $u$ is the specific
thermal energy, and $\gamma = 5/3$ is the ratio of the specific heats
(adiabatic index).  We are assuming that the gaseous disk represents
the partially ionized hydrogen component of the galaxy.  In its
general form the code solves an internal energy equation that includes
an artificial viscosity term to model irreversible heating from
shocks.  The code adopts the standard Monaghan artificial viscosity
and the Balsara criterion to reduce unwanted shear viscosity (Balsara
1995).  In the adiabatic runs the thermal energy can rise as a result
of compressional and shock heating and can drop because of expansion.
In runs including radiative cooling energy can be released also
through radiation. We use a standard cooling function for a primordial
gas composition (helium and atomic hydrogen).

\begin{figure}[!ht]
\plottwo{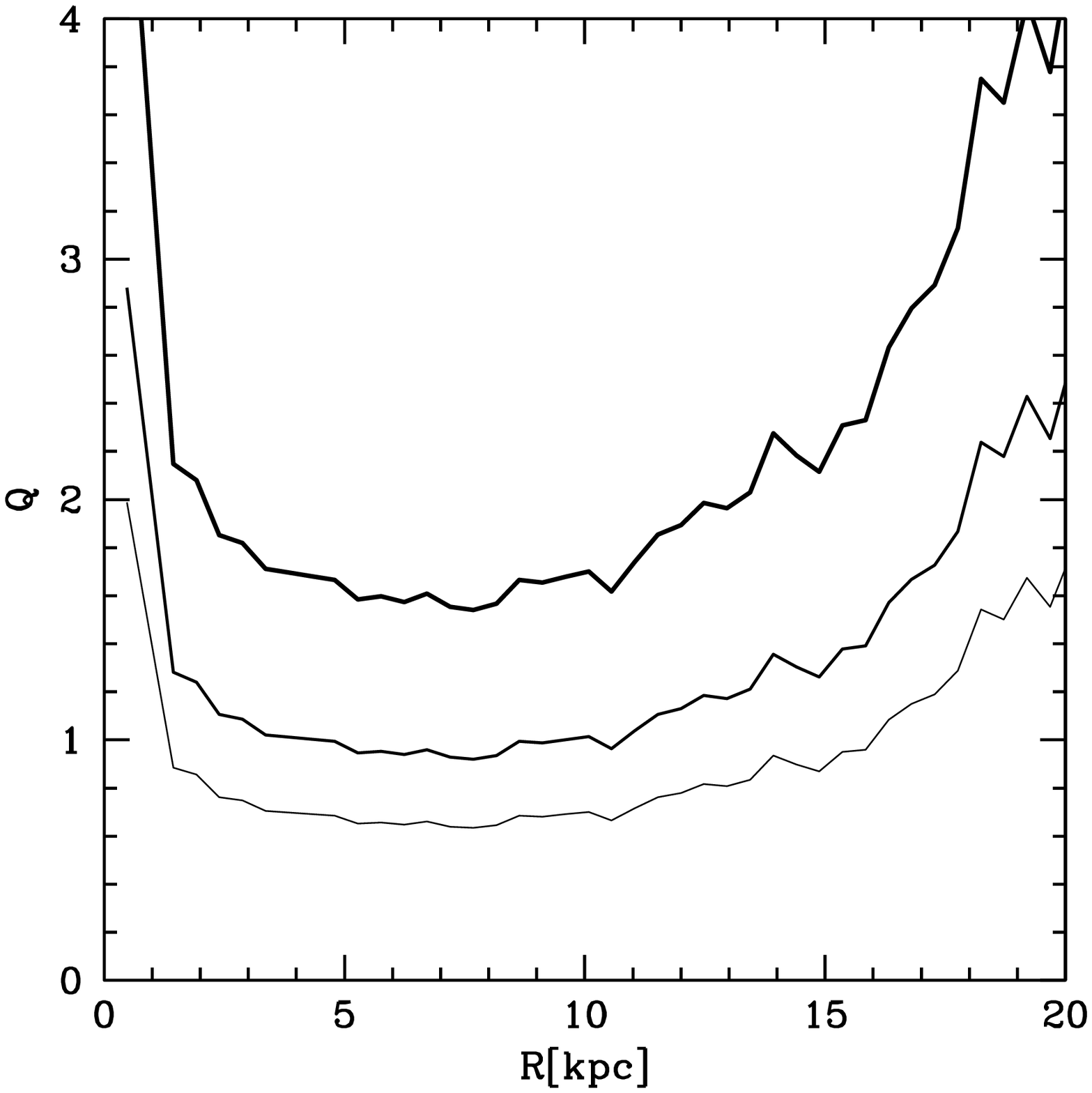}{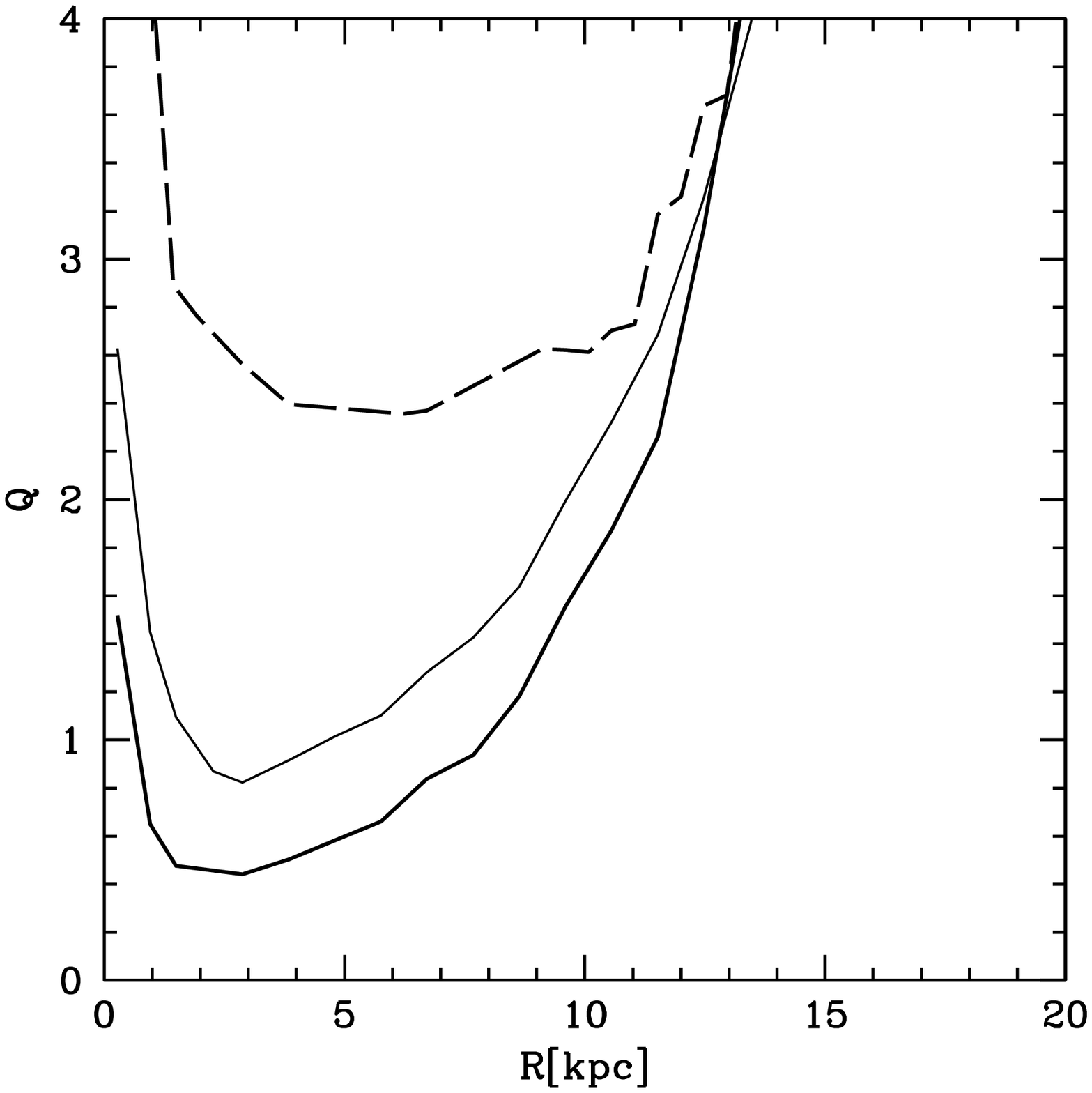}
\caption{Initial Toomre-$Q$ profiles for the collisionless live-halo
 models ({\it left}) and of those including a gaseous component ({\it
 right}). In the left panel we show the profiles of the models used in
 runs NC1, NC2 and NC3 in order of increasing line-thickness. In the
 right panel we show the $Q$ profile of the gaseous component in run
 NG4 ({\it thick solid line}) and run NG5 ({\it thin solid line}). The
 $Q$ profile of the gaseous component in runs NG1-NG3 (not shown) is
 the same as for NG5 except for the normalization being a factor of 5
 higher (see Table \ref{tab:live}).  The $Q$ profile of the stars for
 run NG4 is also shown ({\it thick dashed line}). Stellar $Q$ profiles
 of runs NG1-NG3 and NG5 are equal to that of NC3 except for the
 normalization (see Table \ref{tab:live}).
\label{fig:Qprofs}}
\end{figure}

Dissipational galaxy models are built following the same prescription
described in Section \ref{ssec:livehalos} for live-halo models but
include also a gaseous disk represented by $10^5$ SPH particles each
with a gravitational softening of $300$ pc. The basic properties of the
runs performed are shown in Table \ref{tab:live}. The disk mass
fraction is $f_d=0.05$ in all runs except run NG4, which has
$f_d=0.12$.  The gaseous disk has a temperature of $10^4$ K,
consistent with the gas velocity dispersions derived in observations
(Martin \& Kennicutt 2001).  The gaseous disk has an exponential
surface density profile with the same scale-length as the stellar disk
(see Mayer \& Wadsley 2004), and its thickness is determined by local
hydrostatic equilibrium.  In a gaseous disk the Toomre parameter is
defined as $Q(R)= c_s \kappa/\pi G \Sigma_g$, where $c_s$ is the sound
speed and $\Sigma_g$ is the surface density of the gas. The global
stability of the disk will be determined by the combined stability
properties of the stellar and gaseous disks (Jog \& Solomon 1991).
Gravitational instabilities can be more vigorous in a cold gaseous
disk and might affect the development of non-axisymmetry even in the
stellar disk (Rafikov 2001). In particular, in models having $50\%$
gas, the gaseous disks have $Q<2$ over most of the radial extent of
the galaxy, which should make the system unstable to non-axisymmetric
perturbations irrespective of the stellar $Q$ (see Rafikov 2001 for
the case in which the sound speed $c_s$ is $\sim 0.3$ of the radial
stellar velocity dispersion as in our models).  In the models with
$10\%$ gas instead, the gaseous disk has a high $Q$, making them
stable to axisymmetric perturbations although they still can be
unstable to non-axisymmetric perturbations since $Q < 2$ for the stars
(Binney \& Tremaine 1987).  The $Q$ profiles of gas and stars are
shown in Figure \ref{fig:Qprofs}.

\begin{figure}[!ht]
\plottwo{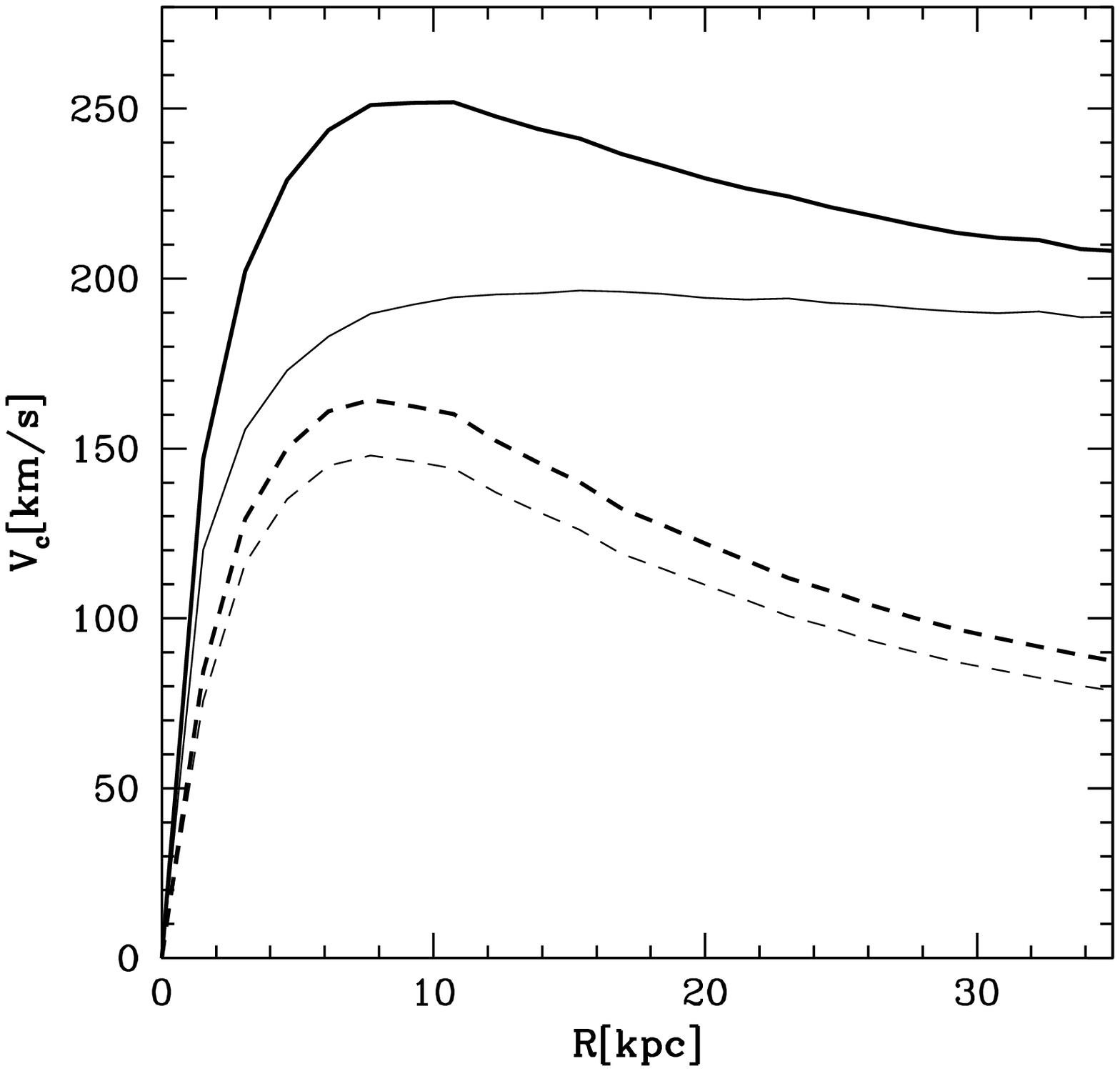}{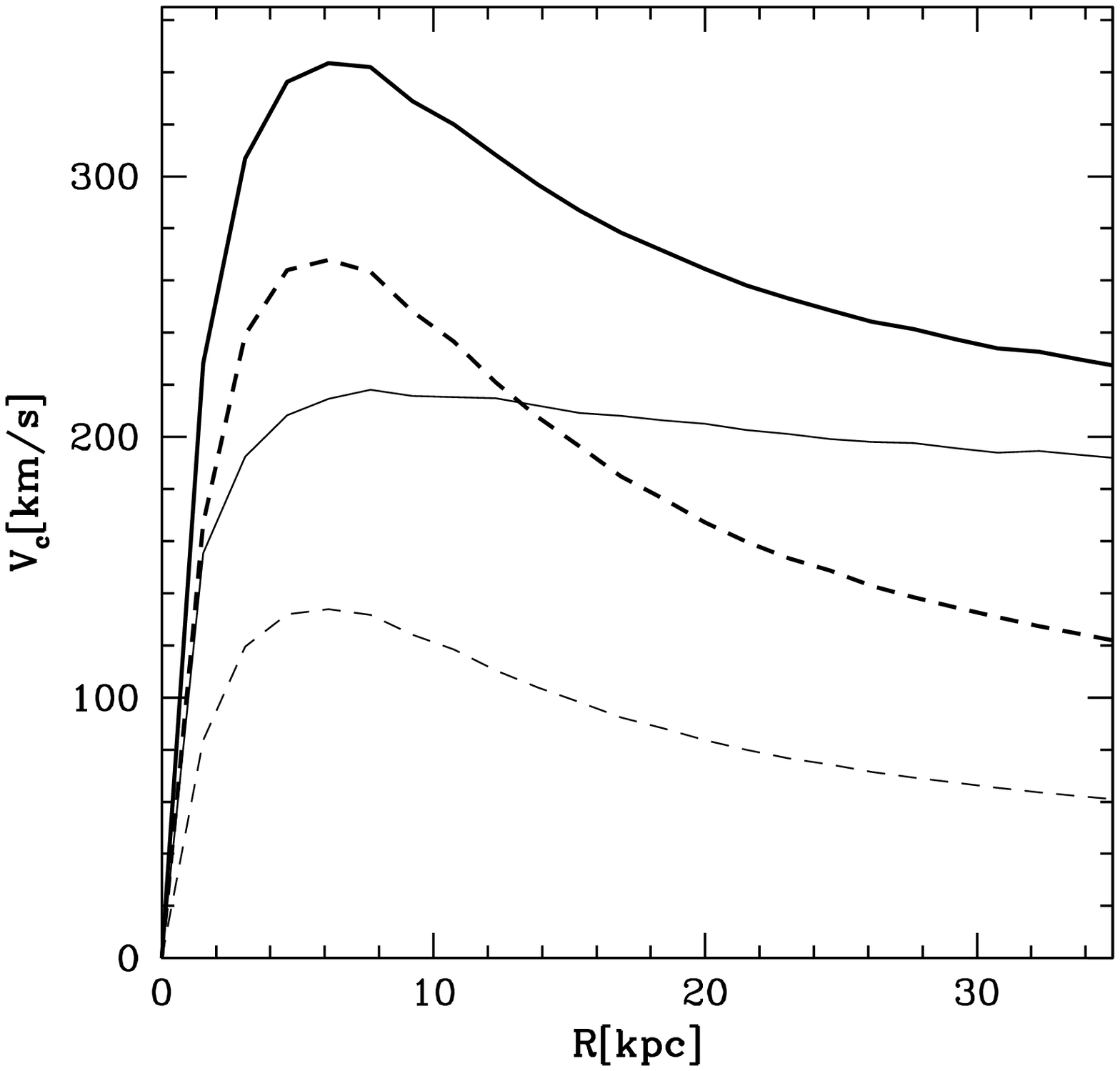}
\caption{Initial rotation curves of the dissipational live-halo
 models. On the left is the mass model used in runs NG1, NG2, and NG3,
 and on the right that used in run NG4.  The thick solid and thick
 dashed lines are, respectively, the total curve and the contribution
 of the baryons (stars+gas), while the thin solid and thin dashed
 lines represent the separate contributions of, respectively, the dark
 matter and stellar component. The rotation curve of run NG5 (not
 shown) is the same as that in the left panel except that the curve
 for gas gets gets shifted up by a factor of $\sim 1.7$ (corresponding
 to a factor of 5 in mass).
\label{fig:rc2}}
\end{figure}

Finally, we also include star formation. The star formation algorithm
follows that of Katz (1992), where stars form from cold,
Jeans-unstable gas particles in regions of convergent flows (see also
Governato \etal\ 2004; Stinson \etal\ 2006). The star formation
efficiency parameter $c_* = 0.15$, but with the adopted scheme its
value has only a minor effect on the star formation rate (Katz 1992).
No supernova feedback is included in our simulations.  The rotation
curve of the models with gas can be seen in Figure \ref{fig:rc2}.

\subsection{Rigid-halo models}

These simulations consist of a live disk inside a rigid halo, which
permit large numbers of particles and thus allow high spatial
resolution to be reached.  High resolution is particularly useful for
studying the vertical evolution of disks.  Rigid-halo simulations are
better suited to systems in which the disk is dominant in the inner
regions because the interaction with the halo is intrinsically
weaker (Debattista \& Sellwood 2000).

The rigid halos were represented by either a logarithmic potential with
a core
\begin{equation}
\Phi_L(r) = \frac{v_{\rm h}^2}{2}~ \ln(r^2 + r_{\rm h}^2),
\end{equation}
or a cuspy Hernquist model
\begin{equation}
\Phi_H(r) = -\frac{M_{\rm h}}{r+r_{\rm h}}.
\end{equation}

The initially axisymmetric disks were all S\'ersic (1968) type,
\begin{equation}
\rho_{\rm d}(R,z) = \frac{M_{\rm d}}{2 \pi R_{\rm d}^2} e^{-(R/R_{\rm
d})^{(1/n)}} \frac{1}{\sqrt{2 \pi} z_{\rm d}}e^{-\frac{1}{2}(z/z_{\rm
d})^2}
\end{equation}
with scale-length $R_{\rm d}$, mass $M_{\rm d}$, and Gaussian
thickness $z_{\rm d}$, truncated at a radius $R_t$ and a S\'ersic
index $n$.  Disk kinematic setup used the epicyclic approximation to
give constant Toomre-$Q$ and the vertical Jeans equation to set
vertical motions appropriate for a constant thickness.  The disks were
represented by $4-7.5\times 10^6$ equal-mass particles.  In units
where $R_{\rm d} = M_{\rm d} = G = 1$, which gives a unit of time
$(R_{\rm d}^3/GM_{\rm d})^{1/2}$, the values for the disk$+$halo
parameters such that our rotation curves were always approximately
flat to large radii are given in Table \ref{tab:rigid}.  One possible
scaling to real units has $R_{\rm d} = 2.5$ \kpc\ and $V_c = 200$
\kms, so that $M_{\rm d} = 2.3 \times 10^{10} M_\odot$ and the unit of
time is 12.4 Myr.  We adopt this time scaling throughout but present
masses, lengths, and velocities in natural units.

These simulations were run on a three-dimensional cylindrical polar
grid code (described in Sellwood \& Valluri 1997) with $N_R\times
N_\phi \times N_z = 60 \times 64 \times 243$.  We also ran tests with
finer grids to verify that our results are not sensitive to the grid
used.  The radial spacing of grid cells increases logarithmically from
the center, with the grid reaching to $\sim 2 R_t$ in most cases;
except where noted, $R_t = 5 \rd$.  For all of the simulations in
Table \ref{tab:rigid}, the vertical spacing of the grid planes,
$\delta z$, was set to $0.0125 \rd$ (except in run L1 where we reduced
this to $0.0083 \rd$), but we confirmed that our results do no change
with smaller $\delta z$.  We used Fourier terms up to $m=8$ in the
potential,\footnote{In order that models remain centered at the origin,
we excluded the $m=1$ term in the expansion.  Including this term
would result in large but artificial offsets between the bar and the
center, similar to the ones found by McMillan \& Dehnen (2005).}
which was softened with the standard Plummer kernel, of softening
length $\epsilon = 0.017 \rd$, although we also tested smaller
$\epsilon$ and larger maximum $m$.  Time integration was performed
with a leapfrog integrator with a fixed time-step, $\delta t = 0.01
(\equiv 1.24 \times 10^5 \mathrm{yr})$ for all runs with $n=1$;
otherwise, we use $\delta t = 0.0025 (\equiv 3.1 \times 10^4
\mathrm{yr})$.  With these values, a circular orbit at $R_{\rm d}/10$
typically is resolved into 600 steps.  For the logarithmic halos, we
set $(r_{\rm h},v_{\rm h}) = (3.3,0.68)$, while the Hernquist halos
had $(r_{\rm h},M_{\rm h}) = (20.8,43.4)$.

\begin{table*}[!ht]
\begin{centering}
\begin{tabular}{cccccc|cccccc}\hline 
\multicolumn{1}{c}{Run} &
\multicolumn{1}{c}{$z_{\rm d}/\rd$} &
\multicolumn{1}{c}{$Q$} &
\multicolumn{1}{c}{$n$} &
\multicolumn{1}{c}{Halo} &
\multicolumn{1}{c}{$r_{\rm h}/\rd$} &
\multicolumn{1}{|c}{$\ln A_\phi$} &
\multicolumn{1}{c}{$\ln (A_z/\rd)$} &
\multicolumn{1}{c}{$B/D$} &
\multicolumn{1}{c}{$n_b$} &
\multicolumn{1}{c}{$R_{b,eff}/R_{d,f}$} &
\multicolumn{1}{c}{$R_{d,f}/\rd$} \\ \hline 
% 292
L1 & 0.025& 1.6 & 1.0 & Log. & 3.3  & -1.54 & -3.80 & 0.44 & 1.1 & 0.32 & 1.7 \\ 
% 240                                                                                                          
L2 & 0.05 & 1.2 & 1.0 & Log. & 3.3  & -1.15 & -3.45 & 0.52 & 1.3 & 0.17 & 2.1 \\
% 241                                                                                                          
L3 & 0.05 & 1.6 & 1.0 & Log. & 3.3  & -1.36 & -3.77 & 0.56 & 0.9 & 0.47 & 1.7 \\
% 252                                                                                                          
L4 & 0.10 & 1.2 & 1.0 & Log. & 3.3  & -1.77 & -8.07 & 0.36 & 0.8 & 0.34 & 1.7 \\
% 253                                                                                                          
L5 & 0.10 & 1.6 & 1.0 & Log. & 3.3  & -1.36 & -7.19 & 0.69 & 0.8 & 0.68 & 1.5 \\
% 264                                                                                                          
L6 & 0.20 & 1.2 & 1.0 & Log. & 3.3  & -1.11 & -8.07 & 0.53 & 0.8 & 0.44 & 1.8 \\
% 303                                                                                                          
S1 & 0.05 & 1.0 & 1.5 & Log. & 3.3  & -0.81 & -3.54 & 0.61 & 1.8 & 0.16 & 2.3 \\ 
% 288                                                                                                          
S2 & 0.05 & 1.0 & 2.0 & Log. & 3.3  & -1.20 & -3.76 & 0.76 & 3.1 & 0.14 & 1.5 \\
% 287                                                                                                          
S3 & 0.05 & 1.0 & 2.5 & Log. & 3.3  & -1.27 & -5.91 & 0.63 & 1.7 & 0.12 & 1.5 \\
% 279                                                                                                          
H1 & 0.05 & 1.2 & 1.0 & Hern.& 20.8 & -1.39 & -5.68 & 0.46 & 1.1 & 0.22 & 2.4 \\
% 280                                                                                                          
H2 & 0.05 & 1.6 & 1.0 & Hern.& 20.8 & -1.54 & -6.83 & 0.37 & 1.1 & 0.51 & 1.4 \\
% 281                                                                                                          
H3 & 0.05 & 2.0 & 1.0 & Hern.& 20.8 & -1.94 & -5.22 & 0.06 & 1.2 & 0.70 & 1.0 \\ \hline
\end{tabular}
\caption{The primary sample of rigid-halo simulations in this paper.}
$z_{\rm d}$, $Q$ and $r_{\rm h}$ are the disk Gaussian scale-height,
Toomre-$Q$ and halo scale-length, respectively.  $n$ is the index of
the initial S\'ersic disk ($n=1$ is an exponential disk).  In column
``Halo'' we describe the halo type: a logarithmic or Hernquist
potential.  $\ln A_\phi$ and $\ln A_z$ are the maximum amplitudes of
the bar and of buckling.  Strong buckling corresponds to $\ln A_z
\gtsim -4$.  The quantities $B/D$, $n_b$, $R_{b,eff}/R_{d,f}$ and
$R_{d,f}$ are all parameters of the bulge$+$disk decomposition at the
end of the simulation.  Here $R_{d,f}$ is the final value of the
scale-length.  The following simulations were presented also in Paper
II: L2 (as R1), S3 (as R4), L6 (as R6) and H2 (as R7).
\label{tab:rigid}
\end{centering}
\end{table*}

\subsection{Tracking structural evolution}

Our models host disks that are massive enough to form bars in a few
dynamical times. The formation of the bar is just one of the
mechanisms that drive the morphological evolution of the disks in our
simulations. Spiral structure and vertical instabilities, like the
buckling instability, also lead to secular evolution of disk
structural parameters, from stellar density profiles and disk
scale-lengths to the bulge-to-disk ratio.
In order to track the evolution of our models, we measured the
amplitude of the bar, $A_\phi$, as the normalized amplitude of the
$m=2$ density distribution:
\begin{equation}
A_\phi = \frac{1}{N} \left| \sum_j e^{2 i \phi_j}\right|
\end{equation}
where $\phi_j$ is the two-dimensional cylindrical polar angle
coordinate of particle $j$.  The sum extends over stellar particles
only.  We measured the $m=2$ bending amplitude, $A_z$, similarly:
\begin{equation}
A_z = \frac{1}{N} \left| \sum_j z_j e^{2 i \phi_j}\right|.
\end{equation}
These quantities allowed us to determine when a bar formed and whether
it buckled.

\section{Vertical evolution}

The vertical direction is best resolved in the rigid-halo simulations,
so we begin by considering those. By far the fastest secular evolution
in the vertical direction is driven by the buckling instability.  This
bending instability, which is caused by anisotropy, is very efficient
at heating the disk vertically.

Raha (1992) described the distortion of a bar during buckling.  As
that work is not widely available, we present a description of
buckling in run L2 before exploring its effects on stellar disks.  In
the animation accompanying this paper (see also Figure
\ref{fig:bending}) we show the evolution of this run between $t=1.0$
and $t=2.2$ Gyr.  At $t=1.12$ Gyr, the system is largely symmetric
about the midplane but develops a small bend by $t=1.18$ Gyr, which
displaces the center toward positive $z$ and the outer parts of the
bar toward negative $z$.  In the outer parts of the bar, where it has
its largest vertical excursion, the bend develops on the leading side
of the bar where it persists for some time, eventually evolving into a
trailing spiral.  As it passes the major axis of the bar, it grows
substantially.  After this bend has dissipated, the process repeats
another two times (see also Martinez-Valpuesta \etal\ 2006), with
small bends on the leading side of the bar, developing into stronger
bends on crossing the bar's major axis.  At smaller radii within the
bar, the peak bending amplitude occurs close to the minor axis (\eg\
at $t=1.43$ Gyr).  The region outside the bar also bends (\eg\ at
$t=1.55$ Gyr) but generally with smaller amplitude.  Small-scale
bending persists to late times and is still ongoing as late as
$t=2.60$ Gyr.  At $t\simeq 1.42$ Gyr, the buckling produces the
largest mean vertical displacement, $\overline{z} \simeq 0.157 \rd$,
which is more than 3 times the initial disk thickness for this
simulation, $z_{\rm d} = 0.05 \rd$.

\begin{figure}[!ht]
%  \plotone{fig4.ps}
\caption{Bending in run L2 at the peak of the bend, $t=1.43$ Gyr in
our adopted scaling.  The online version of this figure links to an
animation of the process.
\label{fig:bending}}
\end{figure}

Buckling leads to a significant vertical heating: at the center
$\sigma_w/\sigma_u$ increases from $\sim 0.4$ to $\sim 0.85$, and
$\sigma_w/\sigma_u$ averaged inside $R=1.2$ increases by a similar
factor, where $\sigma_w$ is the vertical velocity dispersion and
$\sigma_u$ is the radial velocity dispersion.  The three phases of
strong bending (which can be seen in the accompanying animation) can
be identified with three phases of strong vertical heating.

\subsection{Peanuts from buckling}
\label{ssec:peanuts}

After buckling, model L2 becomes distinctly peanut-shaped when viewed
edge-on.  The disk scale-height $h_z$ has increased by factors of
$2-6$ depending on where it is measured.  It is larger on the minor
axis of the bar than on the major, and is smallest at the center,
properties that are typical of all of the rigid-halo simulations but
are more pronounced for the buckled bars (see also Sotnikova \&
Rodionov 2003).  The peanut results in a negative double minimum in
$d_4$, the fourth-order Gauss-Hermite moment (Gerhard 1993; van der
Marel \& Franx 1993) of the density distribution (\ie\ the peanut
produces a flat-topped density distribution) {\it within the bar}.
The peanut is also manifest in the face-on kinematics as a pronounced
negative minimum in the Gauss-Hermite kinematic moment,
$s_4$.\footnote{We use Gauss-Hermite moments in which the velocity
scale is the rms.  In order to distinguish this from the more commonly
used best-fit Gaussian scale, we refer to this moment as $s_4$ rather
than $h_4$.}  No similar signature of a peanut is evident before
buckling.  In Debattista \etal\ (2005; Paper II) we developed this
into a diagnostic signature of peanuts seen nearly face-on.

Buckling need not always result in a peanut.  Both models L1 and L2
formed strong bars and had roughly equal buckling that vertically
heated both disks substantially.  However, whereas L2 formed a strong
peanut, L1 formed only a weak one.

\subsection{Vertical heating with live-halos and gas}

The vertical heating of disks in our simulations with live halos is
dominated by the buckling instability when gas is absent.  When gas is
present, we found that it may suppress buckling, in agreement with
Berentzen \etal\ (1998).  (We tested that this result does not depend
on force resolution by repeating runs with a softening 6 times
smaller, \ie\ 50 pc.)  But this depends on how readily gas dissipates
its thermal energy (see Figure \ref{fig:gasamps}), a point not
appreciated by Berentzen \etal\ (1998) who performed only isothermal
simulations.  When the gas can cool (NG1), buckling is suppressed.  In
this case the bar amplitude decreases significantly because of the
central gas concentration produced in the inflow driven by the bar.
The reduced bar strength implies a reduced radial anisotropy in the
system, which then is less prone to buckling (Berentzen \etal\ 1998).
However, the bars in the $10\%$ gas case are not destroyed (see
Section \ref{sec:barsurvival}), which suggests that the complete
suppression of buckling does not simply reflect the decrease in bar
strength. This is evident especially in run NG3, which has gas cooling
and star formation, in which a fairly strong bar is present and yet
buckling did not occur.  In Berentzen \etal\ (1998) weak buckling was
always associated with weak bars.

\begin{figure}[!ht]
  \plotone{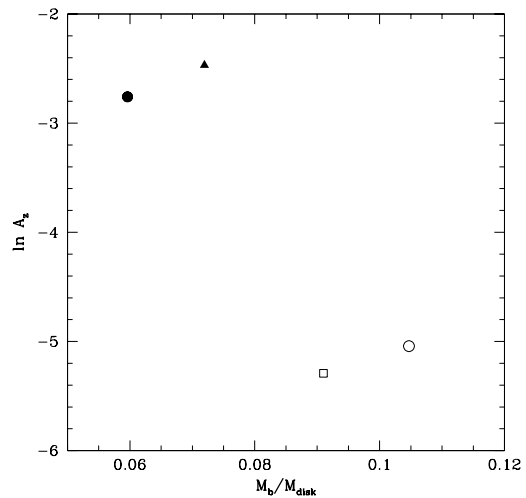}
\caption{Correlation between central mass concentration and the
 maximum strength of the buckling instability in the live-halo
 simulations.  NC3 is indicated by a filled circle, NG2 by a filled
 triangle, NG1 by an open circle, and NG3 by an open square. The
 central mass concentration is measured using the baryonic
 (stellar+gaseous) mass within 800 pc relative to the total disk mass
 (the disk is always the dominant mass component inside 1 kpc in these
 models).
\label{fig:bucklbar}}
\end{figure}

In axisymmetric systems, central concentrations suppress bending modes
(Sotnikova \& Radionov 2005).  Demonstrating a similar result in the
barred case in the presence of gas is non-trivial.  Indeed, whether
gas suppresses buckling directly because it can dissipate bending
energy or because it leads to central mass concentrations (Berentzen
\etal\ 1998) proved difficult to determine because any experiment we
could conceive of also led to different bars.  For example, when we
replaced the central (inner 600 pc) gas blob formed in run NG1 after
$\sim 2$ Gyr with a point mass having equal mass and softening equal
to its half-mass radius and evolved the system as purely
collisionless, we found that the bar buckled, but in the meantime it
also grew stronger.  While we were not able to design a clear test for
these two hypotheses, our results do exhibit a correlation between
buckling amplitude and central mass concentration (Figure
\ref{fig:bucklbar}).

\begin{figure}[!ht]
  \plotone{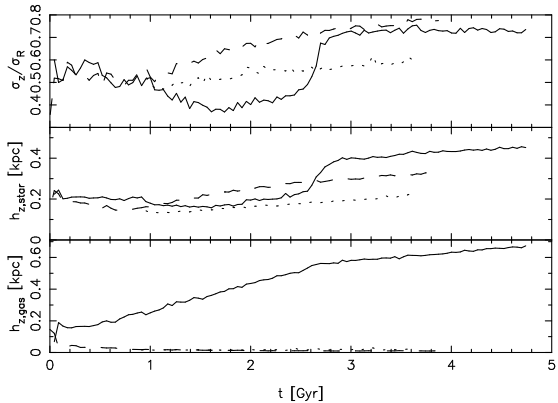}
\caption{Vertical heating in models NG1 ({\it dashed line}), NG2 ({\it
solid line}), and NG3 ({\it dotted line}).  The top panel plots the
ratio of vertical to radial velocity dispersions at the center of the
disks as a proxy for vertical heating.  The middle and bottom panels
show the central scale-heights of stellar and gaseous disks,
respectively.  NG2 buckles, which results in the strong vertical
heating seen at $\sim 2.5$ Gyr, while the bars in NG1 and NG3 never
buckled.  But because their gas disks remain much thinner than the
stellar disks, they continue to heat it vertically over a long time.
\label{fig:gasheating}}
\end{figure}

Vertical heating in the presence of gas occurs even without buckling,
as shown in Figure \ref{fig:gasheating}.  In NG1 and NG3, in which no
buckling occurs, we still see an increase in the vertical stellar
velocity dispersion.  This heating is gentle, with $\sigma_z/\sigma_R$
increasing nearly linearly with time and never falling below the
critical threshold of $\sim 0.4$. The cause of this vertical heating
appears to be scattering by spirals in the gas disk, which remains
significantly thinner than the stellar disk in this simulation.  In
contrast, heating by buckling, as in run NG2 (in which the gas disk
quickly became thicker than the stellar disk) is abrupt.  The gas
thickness in this simulation results from a steadily increasing
temperature as a result of shock heating.

\subsection{Peanuts without buckling}

The live-halo simulations show an alternative way in which peanuts can
form.  The gas-free live-halo simulations all buckled and in the
process formed peanuts no different from those described above.  In
some simulations with gas (\eg\ NG3) we found peanuts forming without
buckling.  It is possible that these formed by direct resonant
trapping of orbits in the growing bar potential (Quillen 2002).
Indeed, in run NG3 $A_\phi$ increased by a factor of about 2 over a
period of $\sim 2$ Gyr (Figure \ref{fig:gasamps}).

\subsection{Peanuts in the presence of gas}

\begin{figure}[!ht]
\epsscale{.80}
  \plotone{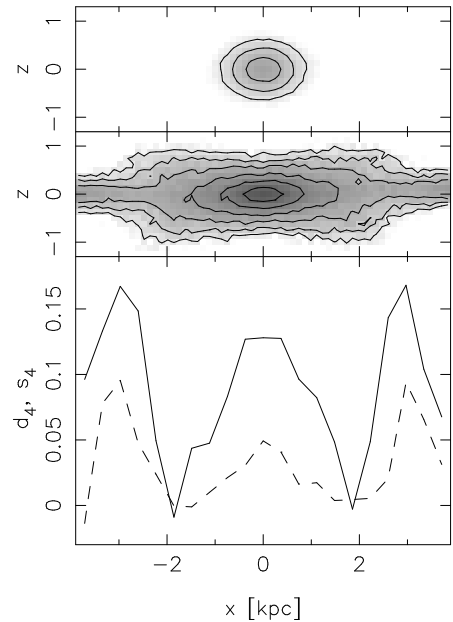}
\caption{Peanut in model NG2.  The top panel shows the edge-on gas
 density in the region $|y| \leq 1$ \kpc, while the middle panel shows
 the stellar edge-on density (in the region $|y| \leq 1$ \kpc).  The
 bottom panel plots $d_4$ ({\it solid line}) and $s_4$ ({\it dashed
 line}).
\label{fig:gaspeanut}}
\end{figure}

We showed in Paper II that peanuts produce prominent minima in the
kinematic Gauss-Hermite moment, $s_4$, when viewed face-on.  In Figure
\ref{fig:gaspeanut} we show that a peanut can still be recognized by a
prominent negative minimum in $s_4$ in model NG2, despite the presence
of gas.  The reason for this is that the gas sinks to smaller radii
than the peanut.

\section{Bar Destruction}
\label{sec:barsurvival}

Both the buckling instability and the growth of massive central
objects have been suggested to destroy bars.  We examine each of these
in this section.

\subsection{The collisionless case}

After Raha \etal\ (1991) showed that buckling weakens bars, it has
often been assumed that bars are destroyed by buckling.  To our
knowledge, that buckling destroys bars has never been demonstrated;
indeed, none of the bars in our simulations were destroyed by
buckling.
Merritt \& Sellwood (1994) noted that buckling grew stronger when
force resolution was increased because then particles have larger
vertical oscillatory frequencies, destabilizing the bar.  Thus, the
most damaging buckling occurs in the rigid-halo simulations, where the
vertical structure is best resolved.  We verified that the survival of
bars after buckling is not due to insufficient resolution by running
an extensive series of numerical tests at higher resolution using
rigid halos.  For these tests, we used model L2, one of the most
strongly buckling simulations.  The bar amplitude $A_\phi$ after
buckling in these tests turned out to not be strongly dependent on any
numerical parameter.
Thus, that buckling does not destroy bars is not an artifact of
insufficient resolution.  This conclusion is also supported by a
higher resolution version of the live-halo simulation of Paper I,
where we increased the number of particles ($N_{halo} = 4\times 10^6$,
$N_{disk} = 2\times 10^5$) and decreased the softening ($\epsilon =
50$ pc for all particles).  Again, although buckling was strong here
as well, the bar was not destroyed.

We also checked that increasing spatial resolution does not lead to
stronger buckling in weakly buckling simulations.  We re-ran
simulations L5 and H1 at higher resolution ($m=32$, $\delta z = 0.005$
and $\epsilon = 0.0083$) and found that $A_\phi$ is barely affected,
demonstrating that vertical frequencies were well resolved and
confirming that the weak bucklings are intrinsic.

\subsection{The effect of bar slowdown}

Although our grid code simulations have high resolutions, they have
rigid halos; thus bar slowdown (Weinberg 1985; Debattista \& Sellwood
1998, 2000; Sellwood \& Debattista 2006) is not included.  Araki
(1985) showed that stability to bending modes in the infinite,
uniform, non-rotating sheet required that $\sigma_w \geq 0.293
\sigma_u$.  As a bar slows, $\sigma_u$ is likely to increase, which
may drive a stable bar to instability.  To test whether this happens,
we slowed down some of our bars by introducing a retarding quadrupole
moment
\begin{equation}
\Phi_{ret} = \Phi_0~ f(R)~ g(s)~ e^{ -2 i (\phi_{\rm bar} - \phi_r)}
\end{equation}
trailing behind a bar.  Here $s = (t-t_0)/(t_1-t_0)$ and $g(s) = -16
s^2 (1 - s)^2$, so that the perturbation is gently switched on at
$t_0$ and off at $t_1$.  The phase of the bar, $\phi_{\rm bar}$, was
computed at each time-step by computing the phase of the $m=2$ Fourier
moment of all of the particles; since the disk also has spirals, there
is a typical uncertainty of order $\pm 15\degrees$ in the bar angle.
We therefore set $\phi_r = 30\degrees$ to be certain that the
retarding potential always trails the bar.  We chose the radial
dependence of the retarding potential to be $f(R) = R/(1+R^2)^2$,
which ensured that it peaks inside the bar radius.

We performed these experiments on runs L2, L4 and H2; a list of all of
the experiments is given in Table \ref{tab:slowruns}.  For run L2, we
switched on the quadrupole shortly after the bar formed and switched
it off before it buckled.  In this case we found that the buckling is
then stronger, which leaves the bar $\sim 20\%$ weaker but still does
not destroy it.

\begin{table}[!ht]
\begin{centering}
\begin{tabular}{cccc}\hline
\multicolumn{1}{c}{Run} &
\multicolumn{1}{c}{$\Phi_0$} &
\multicolumn{1}{c}{$t_0 [Gyr]$} &
\multicolumn{1}{c}{$t_1 [Gyr]$} \\ \hline
L2.s1 & 16. & 0.74 & 1.24 \\ 
L2.s2 & 1.6 & 0.74 & 1.24 \\ 
L4.s1 & 1.6 & 2.60 & 3.10 \\ 
L4.s2 & 4.8 & 2.60 & 3.10 \\ 
H2.s1 & 4.8 & 2.60 & 3.10 \\ 
H2.s2 & 1.6 & 2.60 & 3.10 \\ 
H2.s3 & 0.8 & 2.60 & 3.10 \\ 
H2.s4 & 0.8 & 2.60 & 3.72 \\ \hline
\end{tabular}
\caption{ The series of simulations to test the effect of bar-slowdown
  on the buckling instability.}  $\Phi_0$ measures the relative
  amplitude of the retarding perturbation, while $t_0$ and $t_1$ give
  the time when the perturbation is switched on and off.
\label{tab:slowruns}
\end{centering}
\end{table}

\begin{figure}[!ht]
\epsscale{.80}
  \plotone{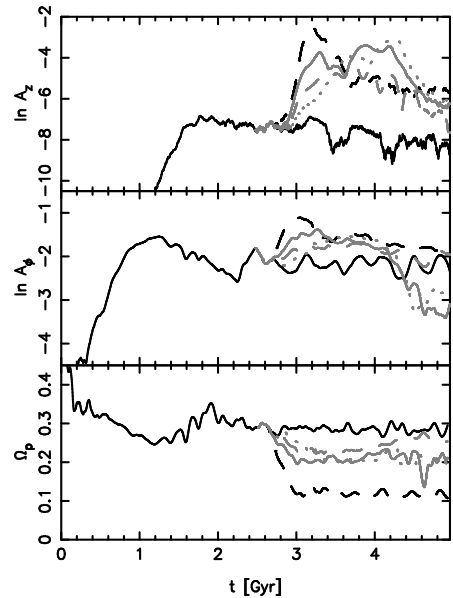}
\caption{Effect of bar slowdown in run H2: whereas the unslowed bar
  suffers only a weak bend, slowdown leads to a fierce buckling.
  However, the bar still survives.  The top and middle panels show the
  buckling and bar amplitudes, respectively, while the bottom panel
  shows the evolution of the bar pattern speed, $\om$.  The black
  solid and dashed lines and the gray solid, dashed, and dotted lines
  show runs H2, H2.s1, H2.s2, H2.s3, and H2.s4, respectively.
\label{fig:slowdown}}
\end{figure}

\begin{figure}[!ht]
  \plotone{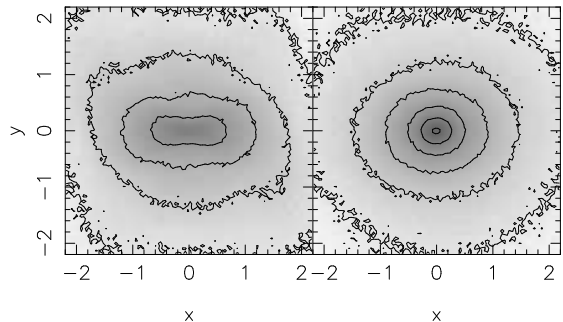}
\caption{On the left is H2.s4 at $t=4.96$ Gyr, showing that before
  buckling the bar was prominent.  On the right the system is shown at
  $t=12.4$ Gyr, well after the imposed bar slowdown.  The bar never
  fully recovers from the strong buckling, and the system is better
  described as weakly barred.
\label{fig:SAB}}
\end{figure}

The other two systems on which we tried such experiments had not
buckled when undisturbed.  In these cases, we turned on the retarding
quadrupole after the bar had settled and turned it off not less than
1.5 bar rotations later.  Slowing down these bars resulted in very
strong buckling, stronger even than in run L2.  But even in these
somewhat extreme cases the bar survives; we illustrate this in Figure
\ref{fig:slowdown}, where we show the various bar slowdown experiments
in model H2.  While these bucklings do not destroy bars, which we
determine simply by visual inspection, in a few cases they leave a
much weaker bar, which would be better described as an SAB than an SB.
In Figure \ref{fig:SAB} we present the most extreme example of H2.s4,
in which the final bar axis ratio was $b/a \simeq 0.85$; although
weak, this can still clearly be recognized by visual inspection.  The
edge-on view of a slice taken around the bar's major axis reveals a
peanut, which can be recognzed by the double minimum in the $s_4$
diagnostic.  These slowed bars probably represent an upper limit to
the damage buckling can inflict on strong bars.

In the live-halo simulations, where the bar can interact with the halo
and slow down, we continue to find that buckling does not destroy
bars.  Therefore we conclude that the bar buckling instability does
not destroy bars (see also Martinez-Valpuesta \& Shlosman 2004).

\subsection{Central gas mass growth}

\begin{figure}[!ht]
\epsscale{.90}
\plotone{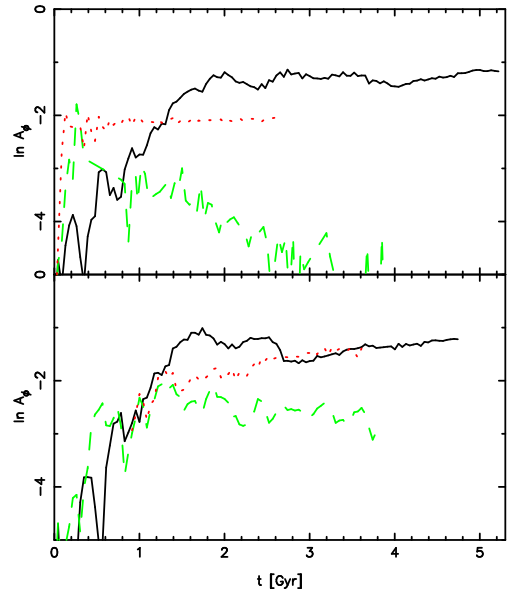}
\caption{Effect of gas on bar amplitudes.  The top panel shows the
  evolution of the bar amplitude in runs NC3 ({\it solid black line}),
  NG4 ({\it dotted red line}), and NG5 ({\it dashed green line}).  The
  bottom panel shows the evolution for runs NG1 ({\it dashed green
  line}), NG2 ({\it solid black line}), and NG3 ({\it dotted red
  line}).
\label{fig:gasamps}}
\end{figure}

\begin{figure}[!ht]
\epsscale{.90}
\plotone{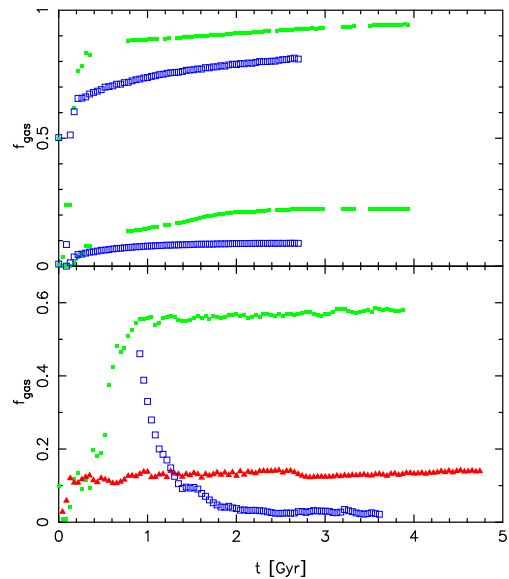}
\caption{Gas mass fraction inside 500 pc in the $10\%$ gas mass runs
  ({\it bottom panel}) and the $50\%$ gas mass runs ({\it top panel}).
  The bottom panel shows NG1 ({\it filled squares}), NG2 ({\it filled
  triangles}), and NG3 ({\it open blue squares}).  The top panel shows
  NG4 ({\it open blue squares}) and NG5 ({\it filled squares}).  The
  lower set of points in this panel show the fraction of the {\it
  total} disk mass, which is inside 500 pc; for the cooling simulation,
  this fraction is large enough to destroy the bar.
\label{fig:fracgas}}
\end{figure}

We now consider bar destruction via the growth of a massive central
gaseous object.  The bottom panel of Figure \ref{fig:gasamps} shows
the evolution of the bar amplitude for models NG1-NG3.  The bar
amplitude depends very strongly on the gas physics: when the gas is
adiabatic (NG2), it does not become very centrally concentrated and
the bar amplitude is not very strongly affected by the gas.  If the
gas can cool (NG1), it sinks quickly to the center and remains there.
Thus, the bar forms already much weaker.  Continued infall at later
times further weakens the bar.  The main difference between the
cooling and adiabatic simulations is in the amount of gas that sinks
into the center of the disk.  In NG1, gas accounts for $\sim 60\%$ of
the mass within 500 pc (Figure \ref{fig:fracgas}); with such a high
fraction, it is unsurprising that the bar is weaker.  The softening in
these simulations is 300 pc; therefore, the central gas mass is not
well resolved.  At these scales the hydrodynamical resolution is
higher than the gravitational force resolution (the gravitational
softening volume contains many times the SPH smoothing volume) with
the result that the collapse of gas toward the center is inhibited
(see Bate \& Burkert 1997). Therefore, the same amount of gas would
probably have collapsed to an even smaller radius had we had greater
force resolution, weakening the bar further.  In NG2 gas cannot
radiate away the intense compressional heating it experiences, and
only $10\%$ by mass is found within 500 pc.  Thus, its bar is stronger
than in NG1.  When star formation is allowed (NG3), the gas
accumulated into the center drives a starburst.  This converts most
of the highly concentrated gas into stars, which can now support the
bar.  A strong bar again forms.  Since our simulations do not include
feedback, none of the mass that falls in flows back out; up to $t\sim
2.5$ Gyr, there is very little difference in the
(azimuthally-averaged) density profile inside 3 kpc between runs NG1
and NG3.  Later profile differences are most likely caused by the
difference in the bars, which, being stronger in run NG3, leads to
further infall to the center.

Not much changes at higher gas mass fraction if the gas is adiabatic
(NG4).  Then the gas layer is quite thick, producing a bar that,
although somewhat weaker than in the purely stellar case (NC3), is
still quite strong.  When the gas can cool (NG5), the gaseous disk
becomes violently gravitationally unstable and a new phenomenon
appears, namely, the fragmentation of gas into clumps that sink to the
center, dragging an associated stellar clump.  Such clump
instabilities have been found in previous simulations (Noguchi 1999;
Immeli \etal\ 2004) and have been shown to build central bulge-like
objects directly.  Immeli \etal\ (2004) showed that the rate at which
clouds dissipate their energy is the main parameter that determines
whether the clump instability occurs.  As a result of the central
mass, only a weak bar forms, and this is eventually destroyed by
continued gas inflow.

In our fully self-consistent simulations with cosmologically motivated
halos, we found that the fraction of the total disk mass needed to
destroy the bar ($\sim 20\%$; Figure \ref{fig:fracgas}) is in very
good agreement with that recently found by Shen \& Sellwood (2004), as
is the gradual decay of the bar amplitude (Figure \ref{fig:gasamps}).
This result is different from that of Bournaud \etal\ (2005); we note
that our model NG5 differs from theirs in two important ways.  NG5 has
a live halo versus their rigid halos, and the dark matter halo is
strongly concentrated at the center.  Both properties of our halos
allow angular momentum to be transferred from bar to halo efficiently
(Weinberg 1985; Debattista \& Sellwood 1998), which may perhaps
account for the difference in these results.

%%%%%%%%%%%%%%%%%%%%%%%%%%%%%%%%%%%%%%%%%%%%%%%%%%%%%%%%%%%%%%%%%%%%%%

\section{Inner Profile Evolution}
\label{sec:innerprofs}

We explore the evolution of density profiles at large radii in \S
\ref{sec:truncations}, and in \S \ref{sec:b+d} we discuss the
bulge$+$disk decompositions that result from profile evolution.  In
this section we explore this evolution qualitatively.

\subsection{Buckling and central densities}

\begin{figure}[!ht]
\epsscale{.80}
  \plotone{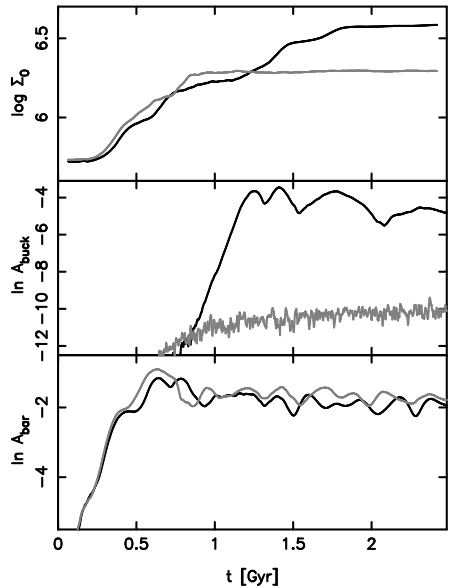}
\caption{Evolution of the central density obtained from particle
  counts within $0.1 R_{\rm d}$ ({\it top}) in the strongly buckling
  run L2 ({\it black line}) and the run with imposed mid-plane
  symmetry ({\it gray line}).  The bottom panel shows the evolution of
  bar amplitude, which reaches similar values in both runs.  The
  middle panel shows the buckling amplitude.  The gray lines show the
  evolution when symmetry about the mid-plane is imposed; while there
  is little difference in bar amplitude, no buckling can occur.
\label{fig:profevol}}
\end{figure}

Bar formation leads to a change in density profiles.  As was already
noted by Hohl (1971), the central density generally increases while
the outer disk becomes shallower.  In Figure \ref{fig:profevol} we
show that bar formation leads to an increase in the central density in
run L2 (here we define central density from particle counts inside
$0.1 R_{\rm d} \simeq 5 \epsilon$).  Moreover, as already noted by Raha
\etal\ (1991), buckling may also increase the central density of a
disk.  We demonstrated {\it directly} that buckling is responsible for
an increase in central density by re-simulating this system with an
imposed symmetry about the mid-plane, which prevents buckling.  Figure
\ref{fig:profevol} shows that when buckling is absent, no further
increase in central density occurs.  The buckled system is 2.0 times
denser at the center than when it is prevented from buckling.

\subsection{Disk scale-lengths and angular momentum}

In Paper I we argued that the fact that the evolution of the
scale-length of the outer disk changes under the influence of the bar,
even when the total baryonic angular momentum is conserved, implies
that the distribution of disk scale-lengths does not follow
automatically from that of halo angular momenta.  The increase in disk
scale-length is due to transfer of angular momentum from the bar to
the outer disk (Hohl 1971).  It is remarkable that the disk outside
the bar remains exponential out to the point where a break occurs,
in both collisionless and dissipative systems.  Hohl (1971) was the
first to notice that exponential disks are naturally obtained after
bar formation (outside the bar) even when the initial disk did not
have an exponential profile.  Since bars are ubiquitous in galaxies at
low and high redshift (Jogee et al. 2004), it follows that the effect
of secular evolution on disk sizes has to be included in any realistic
galaxy formation model.

Not only do bars change disk scale-lengths, but the amount by which
they change varies dramatically depending on the $Q$ profile, even for
(nearly) identical initial angular momentum.  Consider models H1, H2,
and H3.  These all have the same initial conditions other than
Toomre-$Q$, leading to $\ltsim 10\%$ difference in the total baryon
angular momentum.  Nevertheless, the final values of $R_{\rm d}$ range
from $1.0$ to $2.4$ (Table \ref{tab:rigid}).  We conclude that the
direct mapping of halo spins into a distribution of disk scale-lengths
(e.g. Mo, Mao \& White 1998) will not yield correct predictions.

%%%%%%%%%%%%%%%%%%%%%%%%%%%%%%%%%%%%%%%%%%%%%%%%%%%%%%%%%%%%%%%%%%%%%%

\section{Outer Disk Breaks}
\label{sec:truncations}

Disk densities do not always exhibit a single exponential profile.
More typically a sharp break between an inner and outer profile is
evident.  These breaks are often referred to as truncations following
the apparently sharp drop-offs first discovered by van der Kruit
(1979).  Subsequently, van der Kruit \& Searle (1981a,b) fitted sharp
truncations to light profiles at large radii.  However, de Grijs \etal\
(2001) found that truncations occur over a relatively large region,
rather than sharply.  The larger sample of Pohlen (2002) confirmed
this result; he found that truncations are better described by a
double-exponential profile with a break radius.  For his sample of
mostly late-type systems, Pohlen (2002) estimates that the fraction of
disk galaxies with breaks is $\gtsim 79\%$.  Similar disk breaks have
been found up to redshifts of $z\simeq 1$ (P\'erez 2004; Trujillo \&
Pohlen 2005).
What is the origin of these features in the light distribution of
disks?  In Section \ref{sec:innerprofs} we showed that the angular
momentum re-distribution caused by the bar leads to an increased
central density and a shallower density profile outside this.  This
angular momentum redistribution cannot be efficient to arbitrarily
large radii; thus, we may ask whether secular evolution can give rise
to breaks in density profiles.

Studying breaks in $N$-body simulations is numerically challenging
because they generally occur at large radii, where the density of
particles is low.  We therefore ran several high-resolution rigid-halo
simulations with disks extended as far as $R_t = 12\rd$; with
$7.5M$-particles the initial conditions still had $\sim 24K$ particles
at $R \geq 8 \rd$ and $>3K$ at $R \geq 10 \rd$, sufficient to
properly measure the density profile out to the large radii required.
We chose $R_t$ this large in order to ensure that edge-modes (Toomre
1981) do not interfere with other secular effects.  The series of
simulations we used in this study is listed in Table
\ref{tab:truncruns}.

In order to compare our simulations with observations, we used the
double-exponential fitting form of Pohlen (2002).  We only fit
profiles at $2 \rd \leq R \leq 8 \rd$; the lower limit is needed to
avoid the central bulge-like component.  At very large radii, the
surface density barely evolves because of the low self-gravity
(although all of our models were evolved for at least three rotations
at the outermost radius).  Clearly the profile at larger radii
reflects only our initial conditions.  A reasonable transition radius
between the initial profile and the secularly evolved profile occurs
at $R\simeq 8 \rd$, which we use as our upper limit on the
double-exponential fits.  (For the initial pure exponential profile
this is equivalent to $\sim 7$ mag fainter than the center.)

These fits give three dimensionless quantities: $R_{br}/R_{in}$, the
ratio of break radius to inner scale-length, $R_{out}/R_{in}$, the
ratio of outer scale-length to inner scale-length and
$\mu_{0,in}-\mu_{0,out}$, the difference between the central
surface-brightnesses of the two exponential fits, which we compare
with the data of Pohlen \etal\ (2002) and Pohlen \& Trujillo (2006).

\begin{table}[!ht]
\begin{centering}
\begin{tabular}{ccc}\hline
\multicolumn{1}{c}{Run} & 
\multicolumn{1}{c}{$R_t/\rd$} &
\multicolumn{1}{c}{$Q$} \\ \hline
% 310 
L2.t8  &  8 & 1.2   \\
% 364 
L2.t12 & 12 & 1.2   \\
% 366 
L4.t12 & 12 & 1.2   \\ 
% 370 
L5.t12 & 12 & 1.6   \\ 
% 376 
H1.t12 & 12 & 1.2   \\ 
% 375 
H2.t12 & 12 & 1.6   \\ 
% 365 
H3.t12 & 12 & 2.0   \\ 
% 371 
T1.t12 & 12 & 1.2   \\ \hline
\end{tabular}
\caption{ The simulations testing disk breaks.}  The first two
  characters in the name of each simulation reflects the model from
  Table \ref{tab:rigid} on which that simulation is based by extending
  $R_t$ from $5\rd$ to either $8\rd$ or $12\rd$ (number after ``t'' in
  the name of each run).  Model T1.t12 is not based on any in Table
  \ref{tab:rigid}.  It was produced by adding to run L2 a central
  Gaussian: $\Sigma(R) = \Sigma_0 (e^{-R/R_{\rm d}} + 4 e^{-1/2
  (R/0.2)^2})$.
\label{tab:truncruns}
\end{centering}
\end{table}

\subsection{The face-on view}

\begin{figure*}[!ht]
  \plottwo{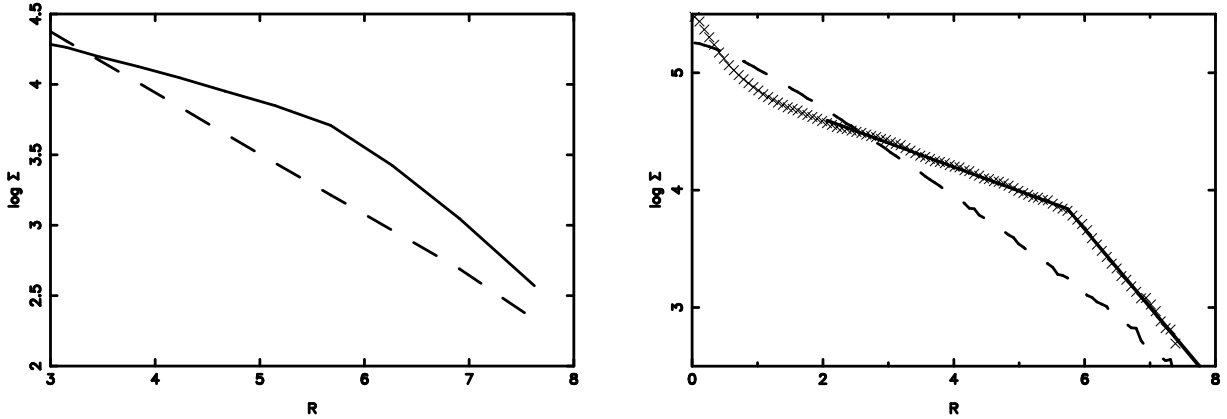}{fig13b.ps}
\caption{Density profile of run L2.t8 at large radii showing the
  formation of a break.  The left panel shows the azimuthally averaged
  face-on density profile, with the dashed line showing the initial
  conditions and the solid line showing the final profile.  In the
  right panel, which shows the edge-on profile, the two solid lines
  indicate the exponential fits over their respective regions, while
  the crosses show the surface brightness.  The dashed line shows the
  edge-on profile of the initial exponential disk.
\label{fig:breakevol}}
\end{figure*}

\begin{figure}[!ht]
\epsscale{.80}
  \plotone{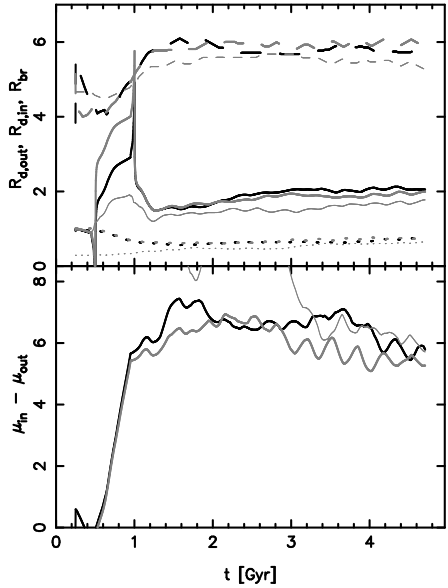}
\caption{Evolution of the break in the face-on surface density.  In
  the top panel we show $R_{out}$ ({\it dotted lines}), $R_{in}$ ({\it
  solid lines}) and $R_{br}$ ({\it dashed lines}).  The black lines
  show run L2.t8 ($R_t = 8\rd$), the thick gray lines show run L2.T12
  ($R_t = 12\rd$), and the thin gray lines show run L2 ($R_t = 5\rd$).
\label{fig:breakrad}}
\end{figure}

Several of our simulations produced clear breaks of the
double-exponential type.  We present one example in Figure
\ref{fig:breakevol}, where we show the initial and final profiles in
run L2.t8.  The profile very quickly develops from a single
exponential to a double-exponential.  In Figure \ref{fig:breakrad} we
plot the evolution of the parameters of the double-exponential profile
for this run and for runs L2 and L2.t12.  The formation of the break
in these three simulations is obviously a discrete event.  Their bars
formed at $t\simeq 620$ Myr; $R_{br}$ does not evolve substantially
after $t\simeq 990$ Myr.  Because of this near coincidence in time, we
conclude that, in these simulations, the process of bar {\it
formation} can directly or indirectly somehow lead to the formation of
broken profiles.

Figure \ref{fig:breakrad} also investigates the difference between
$R_t = 8\rd$ ({\it black lines}), $R_t = 12\rd$ ({\it thick gray
lines}), and $R_t = 5\rd$ ({\it thin gray lines}) (models L2.t8,
L2.t12, and L2, respectively).  The similarity of the break parameters
in the three simulations shows that the breaks do not result from
edge-modes (Toomre 1981).  In run L2, the initial disk did not extend
as far as the final break radius; thus, the break is not an artifact
of our initial disk extending to very large radii.  Therefore, all
parameters of the double-exponential profile that develops can be
considered robust.

\begin{figure}[!ht]
  \plotone{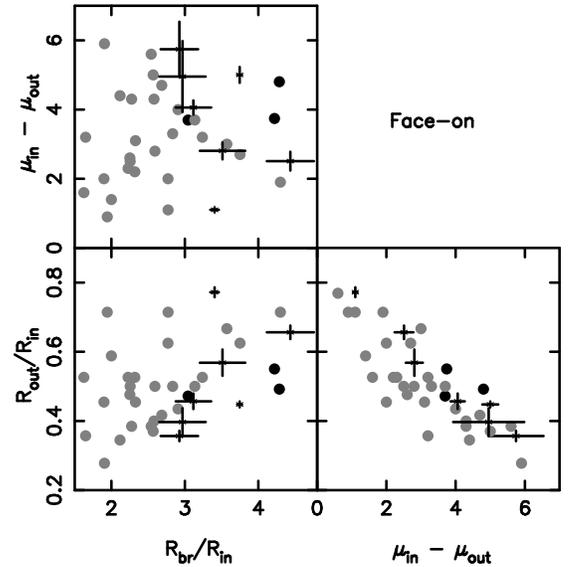}
\caption{Profile break parameters for our simulations seen face-on.
  The filled black circles are data from Pohlen \etal\ (2002), and the
  filled gray circles are data from Pohlen \& Trujillo (2006).  The
  simulation error bars reflect temporal fluctuations; only
  simulations that develop breaks are plotted.
\label{fig:pohlen_faceon}}
\end{figure}

In Figure \ref{fig:pohlen_faceon} we compare the break parameters of
the simulations with the observations of Pohlen \etal\ (2002) and
Pohlen \& Trujillo (2006) for a combined sample of 31 face-on
galaxies.  Our simulations are in reasonable agreement with the
observations, although they span a smaller part of the parameter
space.  The agreement in the narrow distributions in the
$(R_{out}/R_{in}$, $\mu_{0,in}-\mu_{0,out})$-plane is quite striking.

\begin{figure*}[!ht]
\plottwo{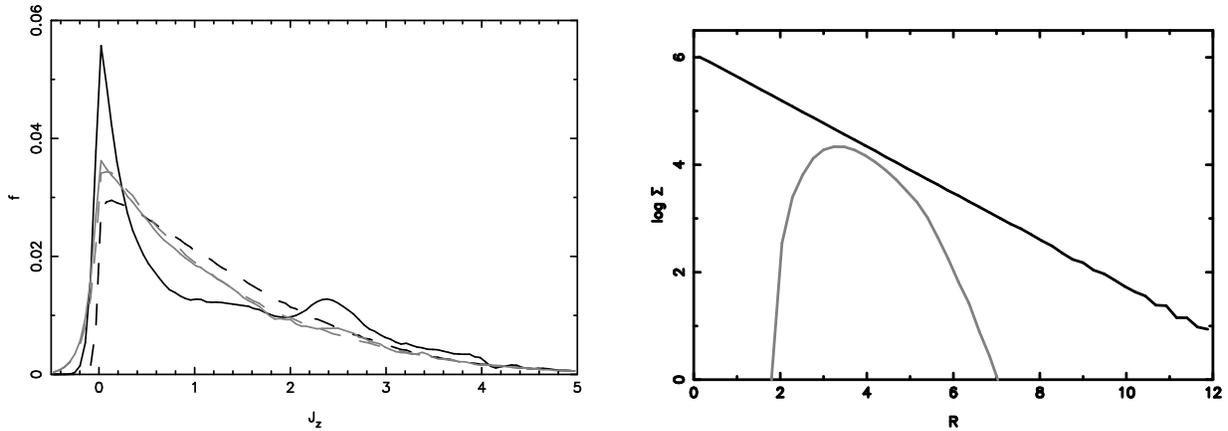}{fig16b.ps}
\caption{{\it Left}: Initial ({\it dashed lines}) and final ({\it
solid lines}) distribution of angular momenta in runs H1.t12 ({\it
black lines}) and H3.t12 ({\it gray lines}).  {\it Right}: Full
initial surface density ({\it black line}) and final surface density
of particles with $2.1 \leq j_z \leq 2.6$ ({\it gray line}) in run
H1.t12.
\label{fig:angmom}}
\end{figure*}

We explore the angular momentum redistribution that leads to the
breaks in Figure \ref{fig:angmom}.  This plots the distribution of
angular momenta in the initial conditions and at the end of the
simulation for runs H1.t12 (with $R_{br} \simeq 5\rd$) and H3.t12
(which did not form a break).  In the left panel we see that very
little angular momentum redistribution occurred in run H3.t12.  On the
other hand, in run H1.t12, bar formation leads to an excess of low
angular momentum particles.  At the same time, a second smaller peak
forms at $2.1 \ltsim j_z \ltsim 2.6$ (see also Pfenniger \& Friedli
1991).  The right panel plots the location of particles in this
angular momentum range.  We find that the bulk of these particles
occur inside the break radius, supporting the interpretation that
breaks occur because of angular momentum redistribution.
 
\subsection{The edge-on view}

Observationally, breaks have often been sought in edge-on systems,
since this orientation leads to higher surface brightnesses.
Comparing to such data is complicated by the fact that these
projections integrate along the entire line-of-sight.  At very large
radii, the density profile does not evolve and reflects initial
conditions.  In order to avoid being biased by these effects, we again
limit our double-exponential fits to {\it projected} radii $R^\prime <
8$; however, we integrate along the entire line-of-sight since to do
otherwise would require an arbitrary cutoff.  In order to increase the
signal-to-noise ratio of our measurements, we use all particles
regardless of their height above or below the disk mid-plane.
Moreover, in the edge-on case, the break parameters depend on the bar
viewing orientation.  Therefore, to compare with simulations, we
consider all orientations of the bar between $0 \leq \phi_{bar} \leq
90 \degrees$.  The right panel of Figure \ref{fig:breakevol} shows an
example of one of our fitted edge-on breaks.

\begin{figure}[!ht]
  \plotone{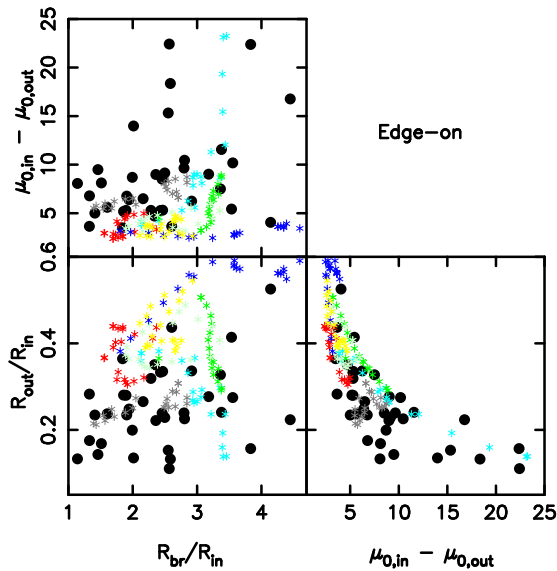}
\caption{ Profile break parameters for edge-on systems.  The filled
  circles are data from Pohlen (2002).  The break parameters are from
  runs L2.t12 ({\it cyan}), L2.t8 ({\it gray}), L4.t12 ({\it green}),
  T1.t12 ({\it red}), H2.t12 ({\it blue}), and H1.t12 ({\it yellow}),
  with varying bar PA also shown.
\label{fig:pohlen_angle}}
\end{figure}

The largest observational sample of disk breaks consists of 37 edge-on
galaxies studied by Pohlen (2002).  We compared our simulations to
these data; the results are shown in Figure \ref{fig:pohlen_angle}.
Variations in $\phi_{bar}$ lead to large variations in the parameters
of the double exponential fit.  Nonetheless, these fall within the
range of observed systems.  This is particularly striking in the
$(\mu_{0,in} - \mu_{0,out},R_{out}/R_{in})$ plane, where the
observational data span a narrow part of the space.  Thus, we conclude
that simple secular evolution suffices to produce realistic disk
breaks.

Figure \ref{fig:pohlen_angle} shows the effect of the line-of-sight
integration: runs L2.t8 ({\it gray asterisks}) and L2.t12 ({\it cyan
asterisks}), which have very similar intrinsic (\ie\ face-on) breaks
(Figure \ref{fig:breakrad}), have very different breaks in the edge-on
view.  Nevertheless, in both cases the resulting parameters are in
good agreement with those in real galaxies.

\subsection{Unbroken profiles}

\begin{figure}[!ht]
  \plotone{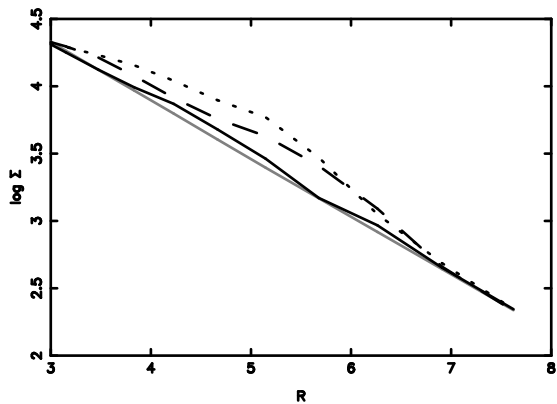}
\caption{Azimuthally averaged surface density in simulations H1.t12,
  H2.t12, and H3.t12, in which only $Q$ of the initial conditions is
  varied: run H1.t12 with $Q=1.2$ ({\it dotted line}), run H2.t12 with
  $Q=1.6$ ({\it dashed line}), and run H3.t12 with $Q=2.0$ ({\it solid
  line}).  The solid gray line shows the initial profile of all three
  simulations.
\label{fig:breakQ}}
\end{figure}

Not all disk galaxies exhibit breaks (Weiner \etal\ 2001; Pohlen \&
Trujillo 2006).  A successful theory of break formation must also be
able to explain such unbroken profiles, which may present difficulties
for a star-formation threshold explanation of disk breaks (Schaye
2004; Elmegreen \& Hunter 2005).  In our simulations, the formation of
a bar is not always accompanied by the formation of a break in the
disk density profile.  For example, run H3.t12 formed a bar, which was
not weak, but the profile remained largely exponential, as can be seen
in Figure \ref{fig:breakQ}.  Similarly, model L5.t12, with $Q = 1.6$,
failed to produce a break.

We investigate the dependence of break parameters on the disk
temperature by considering a set of models in which only $Q$ of the
initial conditions varies (H1.t12, H2.t12, and H3.t12).  We plot the
density profiles in Figure \ref{fig:breakQ}.  At $Q=1.2$, a prominent
break develops in the density profile.  At $Q=1.6$, a break is still
evident, although weaker.  By $Q=2.0$ no break forms in the density
profile.

\subsection{Interpretation: bar-spiral coupling}

The breaks that develop in our simulations are associated with the
angular momentum redistribution induced by the bar.  In all cases the
breaks are well outside the bar semi-major axis (always $\ltsim
3\rd$).  The radius at which the breaks develop instead appears to be
set by spirals resonantly coupled to the bar.  Resonant couplings
between bars and spirals have been described before (\eg\ Tagger
\etal\ 1987; Sygnet \etal\ 1988; Masset \& Tagger 1997; Rautiainen \&
Salo 1999).

\begin{figure*}[!ht]
  \plottwo{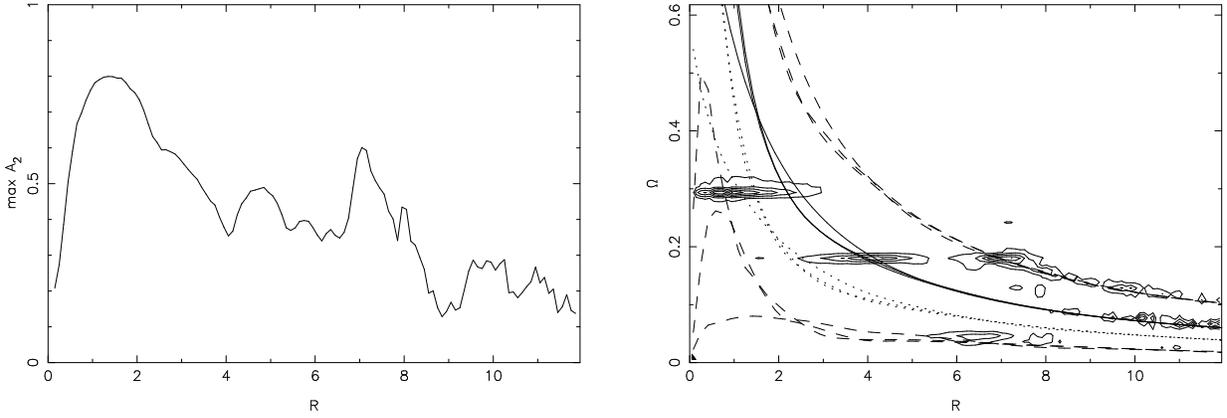}{fig19b.ps}
\caption{{\it Left}: The $m=2$ Fourier peak amplitude of the surface
  density in model L2.t12.  Note the abrupt drop in amplitude at $7\rd
  \leq R \leq 9\rd$.  {\it Right}: Frequencies in run L2.t12.  The
  solid lines show $\Omega(R)$, the frequency of circular rotation.
  The dashed lines show $\Omega-\kappa/2$ ({\it lower}) and
  $\Omega+\kappa/2$ ({\it upper}), while the dotted lines show
  $\Omega-\kappa/4$.  These four sets of lines are plotted at $t=0$,
  $t=1.24$, and $t=4.96$ Gyr (end of the simulation).  The bar forms
  at $t\simeq 0.62$ Gyr.  Contours of the power spectrum are
  overplotted; these show the bar ($\om \simeq 0.29$) and spirals
  ($\om \simeq 0.18$).  This spiral structure appears to be resonantly
  coupled to the bar; its OLR is close to where the break develops in
  the density profile.
\label{fig:radampstrunc}}
\end{figure*}

Evidence for this hypothesis from run L2.t12 is presented in Figure
\ref{fig:radampstrunc}.  The right panel shows the frequency spectrum
of $m=2$ perturbations.  The bar's pattern speed is $\om \simeq 0.29$,
while that of the spirals is $\om \simeq 0.18$.  For the rotation
curve of this system, this puts the corotation radius of the bar at
about the inner 4:1 resonance radius of the spirals.  The outer
Lindblad resonance (OLR) of these spirals is at $R \simeq 7.5\rd$.
The break sets in at $R\simeq 6\rd$, suggesting that it develops
interior to the spiral OLR, at which spiral waves are absorbed.  The
left panel shows that the peak amplitude of $m=2$ perturbations
decreases dramatically between $7\rd \leq R \leq 9\rd$, just outside
where the break develops (see Figure \ref{fig:breakrad}).  We stress
that other spirals with different pattern speeds can and do propagate
to even larger radii.  Resonantly coupled spirals are favored in that
they are stronger, so they transmit more efficiently the angular
momentum shed by the bar during its formation.

For an independent test of this hypothesis we evolved a model with
very different resonance radii: in simulation T1.t12 we forced a
smaller bar with a larger pattern speed by making the disk more
massive in the center (while still exponential in the outer parts).
Assuming a corotation-4:1 bar-spiral coupling, this would bring in the
spiral OLR to roughly $4.8\rd$; indeed we found a break at $R_{br} \simeq
4.5\rd$.  

Thus, we propose that profile breaks develop interior to where the
angular momentum shed by the bar and carried away by
resonantly-coupled spirals is deposited.

%%%%%%%%%%%%%%%%%%%%%%%%%%%%%%%%%%%%%%%%%%%%%%%%%%%%%%%%%%%%%%%%%%%%%%

\section{Bulge+disk decompositions}
\label{sec:b+d}

We compare the simulations with observations of bulges using the
parameters of one-dimensional bulge+disk (bulge+disk) decompositions
and $V_p/\bar{\sigma}$ at a given flattening.  Here $V_p$ is the peak
line-of-sight velocity within some same radial range on the disk
major-axis and $\bar{\sigma}$ is the line-of-sight velocity dispersion
(averaged within the same radial range).

We decomposed the face-on, azimuthally-averaged radial mass profiles
of our simulations into a central S\'ersic and an outer exponential
component, which we will refer to as ``bulge'' and ``disk,''
respectively.  These decompositions are characterized by five
parameters: $\Sigma_{0,d}$, $\Sigma_{0,b}$ (the exponential and
S\'ersic central surface density, respectively), $R_{d,f}$ (final
exponential scale-length), $R_{b,eff}$ (the S\'ersic effective
radius), and $n_b$ (the index of the S\'ersic profile).  In our
bulge+disk fitting, we computed the fits at fixed $n_b$ and obtained
the best fit, including $n_b$, by repeating the fits for $n_b$ in the
range 0.1-4 in steps of 0.1, then selecting the fit with the smallest
$\chi^2$.  We did not distinguish between Freeman types in our
bulge-disk decompositions.  In several cases, therefore, the profile
fits represent a best-fitting {\it average} between small and large
radii.

We compare our simulations with observed galaxies in the dimensionless
space spanned by the parameters $R_{b,eff}/R_{d,f}$, $n_b$, and $B/D$,
the bulge-to-disk mass ratios for profiles extrapolated to infinity.
The photometric data came from two separate studies.  Our first sample
comes from MacArthur \etal\ (2003), who presented bulge+disk
decompositions for a sample of 121 predominantly late-type galaxies of
various inclinations, observed in $B$, $V$, $R$, and $H$ bands.  This
study considered only systems with Freeman type I (\ie\ exponential)
disk profiles in all bands.  As our systems often exhibited transient
type II phases, we also used the decompositions of Graham (2001,
2003).  These were obtained from the diameter-limited sample of 86
low-inclination disk galaxies of all Hubble types observed in the $B$,
$R$, $I$, and $K$ bands by de Jong \& van der Kruit (1994).
It is well known that bulge+disk structural parameters depend on the
filter used (\eg\ M\"ollenhoff 2004).  We compared directly with the
data in all passbands; thus, any discrepancies we find between
simulations and observations are not likely to be due to any
differences in mass-to-light ratios of disks and bulges.

\begin{figure}[!ht]
  \plotone{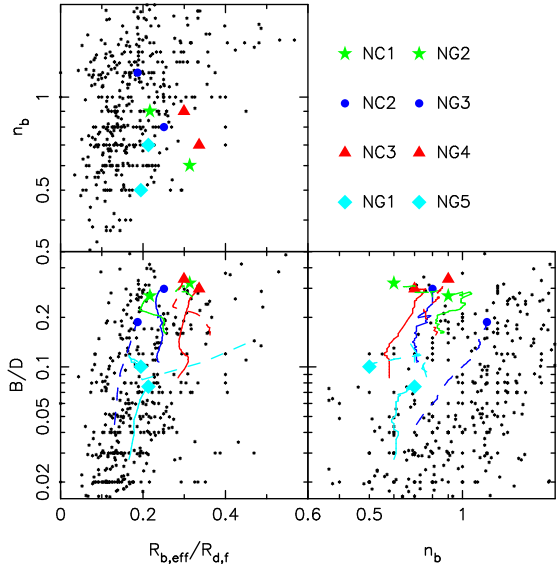}
\caption{Structural parameters of the live-halo runs.  In the
  $(B/D,R_{b,eff}/R_{d,f})$- and $(B/D,n_b)$-planes, the tracks
  indicate the evolution of each simulation {\it after} the bar forms.
  The solid lines refer to runs NC1-NC3 and NG1, while the dashed
  lines refer to runs NG2-NG5.  We do not indicate tracks in the top
  left panel because these are dominated by scatter; the different
  simulations in this panel can be distinguished by comparing the
  ordering of $R_{b,eff}/R_{d,f}$ in the bottom left panel.
\label{fig:livedecs}}
\end{figure}

Paper I presented the decompositions of the rigid-halo models where we
showed that the models partly overlap with the observations but are
mismatched elsewhere.  This mismatch can be diminished if only those
bulges rounder than the disk are considered, but at the cost of
requiring inclinations $i\gtsim 60\degrees$.
In Figure \ref{fig:livedecs} we present the face-on bulge+disk
decompositions for live-halo models after bar formation.  Several
trends are worth noting.  Compared with Paper I, the main changes are
the generally smaller $n_b$ and $B/D$ values, as well as the smaller
discrepancy with observations, although a small discrepancy is still
evident in the $(B/D$,$n_b)$-plane.  All simulations fall in the same
space as observed galaxies in the $(B/D,R_{b,eff}/R_{d,f})$-plane
plane.  In the absence of gas, the values of $B/D$ tend to their
largest values.  Moreover, this quantity increases with time, a result
of the continuing loss of disk angular momentum to the halo, leading
to denser disk centers.  Indeed, we find that the density of the inner
region increases through most of the simulation.  Introducing $10\%$
gas leads to a lower $B/D$ when the gas can cool (NG1) because the
resulting central gas concentration leads to a weaker bar.  When the
gas is adiabatic (NG2), the central mass that grows is significantly
smaller, the bar amplitude is not much different from the
collisionless case, and the $B/D$ ratio is about as large as in the
collisionless systems.  When star formation is included (NG3), the
$B/D$ ratio is intermediate between NG1 and NG2 and is continually
increasing, changing by almost an order of magnitude as gas is turned
into stars.  When gas accounts for $50\%$ of the disk mass, the $B/D$
ratio is again high when gas is adiabatic (NG4) but remains smaller
when it can cool (NG5), in which case a bulge is built via the clump
instability (Noguchi 1999; Immeli \etal\ 2004).

Some overall trends can be noticed.  Generally the systems evolve
parallel to the mean observed correlation between $B/D$ and
$R_{b,eff}/R_{d,f}$ largely because $n_b$ does not evolve much.
In the $(n_b,R_{b,eff}/R_{d,f})$- and $(B/D,n_b)$-planes, the
simulations show a slight tendency to fall outside the observed range,
with $B/D$ evolving toward values larger than observed.  The exception
is model NG3, which evolves parallel to the mean relation between
these two parameters.  This model is also comfortably within the range
of observed bulges in the $(B/D,R_{b,eff}/R_{d,f})$-plane.  Thus,
dissipation with star formation is an important ingredient in the
secular assembly of the bulges seen today.

We compared the kinematics of the bulges with observations in the
$(V/\sigma$,$\epsilon)$-plane at various orientations.  Because of the
smaller number of particles compared with the rigid-halo simulations,
we were not able to fit ellipses as in Paper I.  We therefore obtained
the effective radius of the inclined system through a S\'ersic
bulge$+$exponential disk fit to the mass distribution along the major
(\ie\ inclination) axis.  As in Paper I, because our bulges are poorly
fitted by a de Vaucouleurs profile, we measure kinematic quantities
and mean ellipticities at both one-half and one effective radius and
use the differences as an error estimate.  Ellipticities were measured
from the two-dimensional mass moments of the projected mass
distribution at these two radii, with the difference between the two
giving an error estimate.  Each of the final states of the simulations
is viewed at inclination of $i= 30\degrees$ or $60\degrees$ and for
position angles $\phi_b =0, 45\degrees$, or $90\degrees$.

\begin{figure}[!ht]
  \plotone{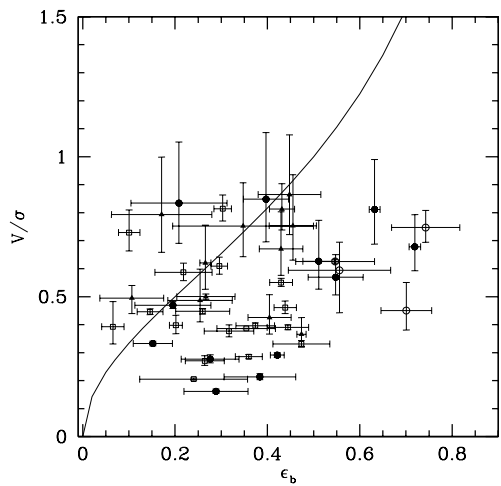}
\caption{Kinematic properties of the live-halo simulations. Open
  squares are for collisionless runs, filled circles are for runs with
  adiabatic gas, filled triangles are for runs with gas and radiative
  cooling, and open circles are for the star formation run. Each run
  was viewed with different inclinations and position angles. Error
  bars are estimated as described in the text.
\label{fig:vsigmanew}}
\end{figure}

In Figure \ref{fig:vsigmanew} we report the results.  The larger
scatter in the live-halo simulations compared with the rigid-halo
simulations (shown in Paper I) is mainly due to the lower number of
particles and related discreteness noise of the mass distribution at
radii only a few times larger than the softening length.  Overall live
and rigid-halo simulations occupy very similar locations, with many
systems below the locus of oblate isotropic systems flattened by
rotation.  Indeed, the simulations include systems significantly
flatter than the observations. These are usually systems in which the
peanut is less pronounced due to weak buckling and the bar still very
strong (a clear example is run NG3), and they are viewed at small
position angles and high inclinations; these objects would appear
markedly bar-like and thus probably would not be included in surveys
of bulges.  Anisotropy clearly plays a role in determining the
flattening, which is unsurprising given that almost all bars
survive. Systems with gas tend to have higher $V/\sigma$ for a given
value of the ellipticity, and the most gas-rich systems are those that
lie closest to the locus of oblate isotropic rotators.  This is
expected if the gas falling toward the center sheds a fraction of its
angular momentum to the stars and spins them up.  Run NG3 produces a
system with the highest flattening due to a combination of a strong
bar and the suppression of the buckling of the bar.

\subsection{The pseudo-bulge formed by the clump instability}

The gas sinking to the center of run NG5 was sufficient to destroy the
bar.  Because we have no star formation in this simulation, we are
left with a rotationally supported massive central gas disk.  Star
formation would have presumably led to a thin, rotationally supported
pseudo-bulge.  The mass of gas within one $R_{b,eff} = 0.83$ kpc
(which is measured face-on from only the stellar component) is $1.2
\times 10^{10} M_\odot$.  The associated aperture dispersion within
$R_{b,eff}$ is $\sigma = 141$ \kms.  In order for the resulting bulge
to sit on the $M_\bullet - \sigma$ relation (Gebhardt \etal\ 2000;
Ferrarese \& Merritt 2000; Tremaine \etal\ 2002), less than $0.3\%$ of
this gas needs to collapse into a black hole. This shows that disk
instabilities are another way, in addition to gas-rich mergers
(Kazantzidis et al. 2005), by which a significant reservoir of gas can
be built to feed an already existing central supermassive black hole
or perhaps produce a new one. Since violent gravitational
instabilities in the gas disk need a high gas mass fraction, such a
mechanism might have played a role in the formation of seed black
holes at high redshift.

%%%%%%%%%%%%%%%%%%%%%%%%%%%%%%%%%%%%%%%%%%%%%%%%%%%%%%%%%%%%%%%%%%%%%%

\section{Discussion and conclusions}

In this paper we explored the secular evolution of disk structural
parameters using simulations.  The initial galaxy model is similar to
a ``Milky Way''-type galaxy that might form within a $\Lambda$CDM
universe.  We have found that bar formation leads to a significant
mass redistribution both in and away from the disk plane. The
evolution of stellar surface density profiles and the formation of
peanut-shaped, dynamically hot stellar structures in the central
regions of the disk are both consequences of bar formation.  Angular
momentum redistribution can lead to large changes in density profiles,
resulting in profiles that can be reasonably fitted by a central
S\'ersic and an outer exponential component.  These can be identified
as bulge and disk components, respectively, although fundamentally
they remain bars.  On purely photometric grounds it is difficult to
distinguish these profiles from bulge+disk decompositions for real
galaxies; however, kinematically these secular bulges clearly fall
below the oblate isotropic rotators.  Real bulges are observed to be
at or above this line.

We have shown that when bars occur in a typical bright spiral galaxy,
they are difficult to destroy.  The buckling instability, which is one
way a peanut-shaped bulge can form, is suppressed when gas is highly
concentrated in the center of the disk.  A peanut can form other than
through a major buckling event in such systems. It is interesting to
note that when buckling occurs, it peaks typically $\sim 2$ Gyr after
the bar forms. It is only at this point that the peanut-shaped bulge
appears.

Our main results can be summarized as follows:

\begin{enumerate}

\item While strong buckling always leads to significant vertical
heating, we found that heating does not always result in a peanut
structure.  When buckling occurs, the central density may increase; in
one simulation this increase more than doubled the central density.
As a consequence, the S\'ersic index of the resulting bulge+disk
decomposition increases somewhat.

\item The effect of gas on the buckling instability depends on the gas
physics.  When the gas evolution included radiative cooling, buckling
was not possible and no peanuts formed.  When the gas is adiabatic,
buckling can occur and peanuts form.  This difference might arise
because the higher central gas concentration suppresses bending modes,
as suggested by the correlation between strength of buckling and
central gas mass concentration. However, a clear test is needed to
discriminate between this scenario and one in which buckling does not
occur because the energy in the modes is dissipated by radiative
cooling. Peanuts can still be recognized by the negative minimum in
the $s_4$ criterion (Debattista \etal\ 2005) even when gas is present
because it sinks to small radii, with the peanut at larger radii.

\item We found no case in which buckling destroyed a bar.  In Paper I
we demonstrated this with the rigid-halo simulations. In this paper
we showed that this result continues to hold when a live halo is
included.  The most damaging buckling events we saw were induced in
the rigid-halo simulations by slowing bars, but even in those cases a
bar survived.  Sometimes, however, the surviving bar is weak and may
be better described as SAB rather than SB.
  
\item Density profiles may evolve substantially under the action of a
bar.  Reasonable bulge+disk decompositions can be fitted to the
resulting profiles.  When comparing the fits with observed galaxies,
purely collisionless secular evolution gives rise to systems
marginally consistent with bulges in nature.  The presence of a modest
($10\%$) amount of gas produces systems that are better able to match
observations.  Star formation helps further and leads to an evolution
of structural parameters parallel to their locus for observed
galaxies.  Secular evolution generally gives rise to nearly
exponential inner profiles.  Kinematically, however, the central
bulge-like components of our simulations clearly fall below the locus
of oblate isotropic rotators (at or above which real bulges occur in
nature), reflecting the fact that they are still, fundamentally, bars.

\item The amount of evolution of a density profile following bar
formation depends sensitively on the Toomre-$Q$ of the initial disk.
When this is small, the inner disk needs to shed a large amount of
angular momentum to form a bar and the central density steepens
considerably.  When the disk is hotter, the density change is smaller.
As a result, the exponential scale-length of the disk outside the bar
region depends on the initial disk kinematics.  Thus, a distribution of
dark matter halo specific angular momenta cannot trivially be related
to a distribution of disk scale-lengths, as is often assumed.

\item Angular momentum redistribution also leads to realistic breaks
in the surface density of disks.  The radius at which breaks occur is
interior to the outer Lindblad resonance of spirals resonantly coupled
to the bar.  When the initial disk is hot, little angular momentum
redistribution occurs and no density breaks occur; thus, secular
evolution can also account for galaxies that do not exhibit any
breaks.  The breaks that result in these simulations are in very good
agreement with observations, including not only the break radii in
units of inner disk scale length but also outer scale-lengths and the
difference between central surface brightnesses of the two
exponentials.
On the other hand, we cannot exclude that angular momentum
redistribution driven by other than bars does not account for some or
most of the observed breaks.  In particular, spirals excited by
interactions may constitute another channel by which such breaks may
form; the presence of breaks in unbarred galaxies may require such a
mechanism.  Since our disk break formation simulations were all
collisionless, we cannot address whether star formation thresholds play
any role in break formation, but two results here suggest that these
may not play a prominent role.  First, we are able to produce disks
without truncations, which models invoking star formation thresholds
may have difficulty in producing.  Secondly, it is clear that the
breaks that do form in our simulations are quite insensitive to extent
of the initial disks.
The ease with which angular momentum redistribution gives rise to
realistic profile breaks, together with the ability to produce also
profiles without breaks, provides strong incentives for exploring such
models further.  Moreover, the type of angular momentum exchange we
are advocating here as leading to breaks need not be necessarily
driven by bars and external perturbations can also play a role.  In
that case, depending on the frequency of the perturbation,
anti-truncations (Erwin \etal\ 2005) may also be possible in a unified
picture.
  
\end{enumerate}

The fraction of disks with bars is $\sim 30\%$ at both low (Sellwood
\& Wilkinson 1993) and high (Jogee \etal\ 2004) redshift, and this
increases to $\sim 70 \%$ at low redshift when measured via
dust-penetrating infra-red observations (Knapen 1999; Eskridge \etal\
2000). Since, as we show, bars are long-lived, we are forced to the
conclusion that disk galaxies know at an early epoch whether or not
they will form a bar and there is simply little room for continued bar
formation as a function of cosmic time.  Thus, secular evolution has
had a long time to act on galaxies.

The results of this paper demonstrate a strong coupling between the
properties and evolution of disk galaxies and their associated inner
and outer morphological and structural parameters.  In contrast,
semi-analytic models of disk structural parameters that invoke
specific angular momentum conservation miss the important effects of
bar formation on disk structure.  As noted in Paper I, a more nuanced
analysis that takes into account secular evolution may help to
alleviate discrepancies between predictions for disk galaxy structure
from cosmological models (\eg\ Mo \etal\ 1998) and observations (De
Jong \& Lacey 2000).

Nonetheless, secular evolution cannot account for all discrepancies
between theory and observation.  An example is the difficulty
cosmological models have in forming bulgeless disks as extended as
those observed (D'Onghia \& Burkert 2004, hereafter DB04).
The $N$-body simulations of DB04 showed that halos with a quiet
merging history since $z = 3$ (which are expected to lead to bulgeless
disks, but see also Springel \& Hernquist [2005]) have a median
$\lambda \simeq 0.023$ with a scatter $\sigma_{\ln \lambda} \simeq
0.3$.  On the other hand, van den Bosch \etal\ (2001) measured their
observed mean at $\lambda \simeq 0.067$.  Thus, the mean of $\lambda$
in the cosmological simulations is more than $3\sigma$ smaller than
observed.
Our simulations indicate that the fraction of galaxies that form pure
exponential disks must be higher (and extends to higher mass galaxies
such as our own Milky Way) before secular evolution turns them into
more disk$+$bulge-like systems.  However, even if all of these systems
would have lower $\lambda$-valuess, the distribution of
$\lambda$-values of quiet merger halos makes it unlikely that secular
evolution can account for the discrepancy between predictions and
observations.  An interesting test of $\Lambda$CDM will be whether
enough halos with quiet merger histories form to account for a higher
fraction of bulgeless galaxies.

\acknowledgments

Discussions with St\'ephane Courteau, Aaron Dutton, Peter Erwin, Lauren
MacArthur, Michael Pohlen, Juntai Shen, and especially Frank van den
Bosch were useful.  Thanks also to Michael Pohlen for sharing his data
with us in advance of publication.  T.Q. and V.P.D. acknowledge
partial support from NSF ITR grant PHY-0205413.  We thank the
anonymous referee for a very careful reading and a detailed report
that helped improve this paper.

\end{document}